\documentclass[11pt,preprint]{aastex}

\usepackage{lineno}
\usepackage{amsmath}
\linenumbers

\slugcomment{Submitted to ApJ}

\shorttitle{Eight $\gamma$-ray Pulsars in Blind Searches}
\shortauthors{Abdo et al.}

\begin{document}

\title{Eight $\gamma$-ray pulsars discovered in blind frequency searches of {\it Fermi} LAT data}
\author{P.~M.~Saz~Parkinson\altaffilmark{1,2}, 
M.~Dormody\altaffilmark{1,2}, 
M.~Ziegler\altaffilmark{1,2},
P.~S.~Ray\altaffilmark{1,3}, 
A.~A.~Abdo\altaffilmark{3,4}, 
J.~Ballet\altaffilmark{5}, 
M.~G.~Baring\altaffilmark{6}, 
A.~Belfiore\altaffilmark{2,7,8},
T.~H.~Burnett\altaffilmark{9}, 
G.~A.~Caliandro\altaffilmark{10}, 
F.~Camilo\altaffilmark{11}, 
P.~A.~Caraveo\altaffilmark{7}, 
A.~de~Luca\altaffilmark{12}, 
E.~C.~Ferrara\altaffilmark{13}, 
P.~C.~C.~Freire\altaffilmark{14}, 
J.~E.~Grove\altaffilmark{3}, 
C.~Gwon\altaffilmark{3}, 
A.~K.~Harding\altaffilmark{13}, 
R.~P.~Johnson\altaffilmark{2}, 
T.~J.~Johnson\altaffilmark{13,15}, 
S.~Johnston\altaffilmark{16}, 
M.~Keith\altaffilmark{16}, 
M.~Kerr\altaffilmark{17}, 
J.~Kn\"odlseder\altaffilmark{18}, 
A.~Makeev\altaffilmark{3,19}, 
M.~Marelli\altaffilmark{7,20}, 
P.~F.~Michelson\altaffilmark{21}, 
D.~Parent\altaffilmark{3,16,22,23}, 
S.~M.~Ransom\altaffilmark{24}, 
O.~Reimer\altaffilmark{25,21}, 
R.~W.~Romani\altaffilmark{21}, 
D.~A.~Smith\altaffilmark{22,23}, 
D.~J.~Thompson\altaffilmark{13}, 
K.~Watters\altaffilmark{21}, 
P.~Weltevrede\altaffilmark{26}, 
M.~T.~Wolff\altaffilmark{3}, 
and K.~S.~Wood\altaffilmark{3}
}
\altaffiltext{1}{Corresponding authors: P.~M.~Saz~Parkinson, pablo@scipp.ucsc.edu; M.~Dormody, mdormody@ucsc.edu; M.~Ziegler, ziegler@scipp.ucsc.edu; P.~S.~Ray, Paul.Ray@nrl.navy.mil.}
\altaffiltext{2}{Santa Cruz Institute for Particle Physics, Department of Physics, University of California at Santa Cruz, Santa Cruz, CA 95064, USA}
\altaffiltext{3}{Space Science Division, Naval Research Laboratory, Washington, DC 20375, USA}
\altaffiltext{4}{National Research Council Research Associate, National Academy of Sciences, Washington, DC 20001, USA}
\altaffiltext{5}{Laboratoire AIM, CEA-IRFU/CNRS/Universit\'e Paris Diderot, Service d'Astrophysique, CEA Saclay, 91191 Gif sur Yvette, France}
\altaffiltext{6}{Rice University, Department of Physics and Astronomy, MS-108, P. O. Box 1892, Houston, TX 77251, USA}
\altaffiltext{7}{INAF-Istituto di Astrofisica Spaziale e Fisica Cosmica, I-20133 Milano, Italy}
\altaffiltext{8}{Universit\'{a} di Pavia, Dipartimento di Fisica Teorica e Nucleare (DFNT), I-27100 Pavia, Italy}
\altaffiltext{9}{Department of Physics, University of Washington, Seattle, WA 98195-1560, USA}
\altaffiltext{10}{Institut de Ciencies de l'Espai (IEEC-CSIC), Campus UAB, 08193 Barcelona, Spain}
\altaffiltext{11}{Columbia Astrophysics Laboratory, Columbia University, New York, NY 10027, USA}
\altaffiltext{12}{Istituto Universitario di Studi Superiori (IUSS), I-27100 Pavia, Italy}
\altaffiltext{13}{NASA Goddard Space Flight Center, Greenbelt, MD 20771, USA}
\altaffiltext{14}{Max-Planck-Institut f\"ur Radioastronomie, Auf dem H\"ugel 69, 53121 Bonn, Germany}
\altaffiltext{15}{Department of Physics and Department of Astronomy, University of Maryland, College Park, MD 20742, USA}
\altaffiltext{16}{Australia Telescope National Facility, CSIRO, Epping NSW 1710, Australia}
\altaffiltext{17}{Department of Physics, University of Washington, Seattle, WA 98195-1560, USA}
\altaffiltext{18}{Centre d'\'Etude Spatiale des Rayonnements, CNRS/UPS, BP 44346, F-30128 Toulouse Cedex 4, France}
\altaffiltext{19}{George Mason University, Fairfax, VA 22030, USA}
\altaffiltext{20}{Universit\'{a} degli Studi dell'Insubria, Via Ravasi 2, 21100 Varese, Italy}
\altaffiltext{21}{W. W. Hansen Experimental Physics Laboratory, Kavli Institute for Particle Astrophysics and Cosmology, Department of Physics and SLAC National Accelerator Laboratory, Stanford University, Stanford, CA 94305, USA}
\altaffiltext{22}{CNRS/IN2P3, Centre d'\'Etudes Nucl\'eaires Bordeaux Gradignan, UMR 5797, Gradignan, 33175, France}
\altaffiltext{23}{Universit\'e de Bordeaux, Centre d'\'Etudes Nucl\'eaires Bordeaux Gradignan, UMR 5797, Gradignan, 33175, France}
\altaffiltext{24}{National Radio Astronomy Observatory (NRAO), Charlottesville, VA 22903, USA}
\altaffiltext{25}{Institut f\"ur Astro- und Teilchenphysik and Institut f\"ur Theoretische Physik, Leopold-Franzens-Universit\"at Innsbruck, A-6020 Innsbruck, Austria}
\altaffiltext{26}{Jodrell Bank Centre for Astrophysics, School of Physics and Astronomy, The University of Manchester, M13 9PL, UK}

\begin{abstract}
We report the discovery of eight $\gamma$-ray pulsars in blind frequency searches using 
the Large Area Telescope (LAT), onboard the \textit{Fermi Gamma-ray Space Telescope}. 
Five of the eight pulsars are young ($\tau_c < 100$ kyr), energetic ($\dot E \ge 
10^{36}$ erg s$^{-1}$), and located within the Galactic plane ($|b|<3^\circ$). The 
remaining three are older, less energetic, and located off the plane. Five pulsars are associated 
with sources included in the \textit{Fermi}-LAT bright $\gamma$-ray source list, but only one, PSR\,J1413$-$6205, is clearly 
associated with an EGRET source. PSR\,J1023$-$5746 has the 
smallest characteristic age ($\tau_c = 4.6$ kyr) and is the most energetic ($\dot E = 1.1 \times 10^{37}$ erg s$^{-1}$) of all 
$\gamma$-ray pulsars discovered so far in blind searches. PSRs\,J1957+5033 and J2055+25 have 
the largest characteristic ages ($\tau_c \sim1$ Myr) and are the least energetic ($\dot E 
\sim5\times10^{33}$ erg s$^{-1}$) of the newly-discovered pulsars. We present the 
timing models, light curves, and detailed spectral parameters of the new pulsars. 
We used recent {\it XMM} observations to identify the counterpart of PSR\,J2055+25 as XMMU\,J205549.4+253959. In addition, 
publicly available archival {\it Chandra} X-ray data allowed us to identify the likely counterpart of 
PSR\,J1023$-$5746 as a faint, highly absorbed source, CXOU\,J102302.8--574606. The large X-ray absorption indicates that 
this could be among the most distant $\gamma$-ray pulsars detected so far. PSR\,J1023$-$5746 is positionally coincident with the 
TeV source HESS\,J1023$-$575, located near the young stellar cluster Westerlund 2, while PSR\,J1954+2836 is coincident with a 
4.3$\sigma$ excess reported by Milagro at a median energy of 35 TeV.
Deep radio follow-up observations of the eight pulsars resulted in no detections of pulsations and upper limits 
comparable to the faintest known radio pulsars, indicating that these pulsars can be included among the 
growing population of radio-quiet pulsars in our Galaxy being uncovered by the LAT, and currently numbering more than 20. 
\end{abstract}

\keywords{gamma rays: general -- pulsars: general -- pulsars: individual (PSR\,J1023--5746, PSR\,J2055+25) -- X-rays: individual (CXOU\,J102302.8--574606, XMMU\,J205549.4+253959) --
open clusters and associations: individual (Westerlund 2)}

\section{Introduction}

Until the launch of the {\it Fermi Gamma-ray Space Telescope}, only seven pulsars were found to have
pulsed emission in $\gamma$ rays~\citep{2008RPPh...71k6901T},
and only one of these pulsars (Geminga) was undetectable by radio
telescopes (i.e. radio quiet). The Large Area Telescope (LAT) has detected
$\gamma$-ray emission from a large number of pulsars using radio
ephemerides, both from young isolated pulsars
\citep[e.g.][]{LATPSR1028,LATPSR2021}, and from millisecond
pulsars \citep{LATMSPs}. In addition, sixteen pulsars were found in
blind frequency searches of the LAT $\gamma$-ray data
\citep{LATBSPs}. Radio follow-up observations of those pulsars resulted in the detection 
of radio pulsations from three of them: PSRs\,J1741--2054 and J2032+4127~\citep{camilo09} and PSR\,J1907+0602~\citep{MGRO}. Two of these were found 
to have remarkably low radio luminosities. Perhaps more surprising is the fact that 13 out of these 16 pulsars were not detected in radio~\citep{TimingPaper}, and 
should therefore be considered radio quiet, or at least radio faint. 

This paper reports on eight new $\gamma$-ray pulsars discovered in blind frequency searches of LAT data, not included in the recently published First 
{\it Fermi} LAT Catalog of $\gamma$-ray Pulsars~\citep{LATPulsars}. As in the pulsar catalog paper, we present detailed timing and spectral results for each of these pulsars, including the pulse 
shape parameters, the fluxes, and the spectral indices and energy cutoffs for each pulsar. We also discuss possible associations, including 
previous $\gamma$-ray detections. While a large number of the first sixteen $\gamma$-ray pulsars discovered in blind searches~\citep{LATBSPs} were found to be 
coincident with previously known $\gamma$-ray sources, such as EGRET unidentified sources, none of the pulsars in this 
sample have a definite counterpart in the 3rd EGRET catalog~\citep{3EGcatalog}, although one pulsar, PSR\,J1413$-$6205, has a
counterpart, EGR\,J1414$-$6244, in a revised catalog of EGRET sources~\citep{EGRcatalog}. This pulsar is also coincident with the AGILE source
1AGL\,J1412$-$6149~\citep{AGILEcatalog}. We also discuss possible X-ray and TeV associations with the newly-discovered pulsars. The 8 pulsars identified here 
are among the 1451 sources included in the {\it Fermi} LAT First Source Catalog\footnote{Available at {\tt http://fermi.gsfc.nasa.gov/ssc/data/access/lat/1yr\_catalog/}}~\citep{LATCatalog}. 

Finally, deep radio follow-up observations of these eight new pulsars have resulted in no new detections of pulsations, and we include the 
upper limits of our radio searches.

\section{Observations and Data Analysis}

The LAT is a high-energy $\gamma$-ray telescope onboard the \textit{Fermi} satellite that is sensitive to photons with energies 
from 20 MeV to over 300 GeV. It features a solid-state
silicon tracker, a cesium-iodide calorimeter and an anti-coincidence
detector~\citep{LATinstrument}. Events recorded with the LAT have
time stamps derived from a GPS-synchronized clock on the \emph{Fermi} satellite
with $<1 \mu$s accuracy \citep{LATcalib}.

For pulsar science, the LAT represents a major advance over EGRET. It has a much larger effective area ($\sim$8000 cm$^2$ for 1 GeV photons at normal 
incidence, or approximately 6 times that of EGRET), a larger field of view ($\sim$2.4 sr, or almost five times that of EGRET), and a finer point spread 
function (68\% containment angle of 0\fdg6 at 1 GeV for the {\it Front} section and about a factor of 2 larger for the {\it Back} section, 
vs. $\sim$1\fdg7 at 1 GeV for EGRET). The LAT also has a more efficient viewing strategy, operating primarily in continuous sky survey mode, as opposed 
to the inertial pointing mode used by EGRET~\footnote{For a single region in particular, however, a pointed observation would be more efficient 
for detecting pulsations, by accumulating more photons over a shorter time period.}. This optimizes the amount of time the sky is in the field of view, covering 
the entire sky every two orbits ($\sim$3 hr). These improvements allow more photons to be accumulated per unit time, and provide a better signal-to-noise 
ratio in photon selection, making the LAT the first highly effective instrument for blind searches of $\gamma$-ray pulsars~\citep{LATBSPs}. For more details 
on the LAT, see \cite{LATinstrument}.

\subsection{Blind Frequency Searches\label{bfs}}

Since $\gamma$-ray photon data are extremely sparse (a typical $\gamma$-ray pulsar flux in the LAT may be of order 1000 photons per year), 
searches for $\gamma$-ray pulsars require long integration times. Furthermore, young rotation-powered pulsars spin down as they radiate 
away energy, so their signals are not precisely periodic, making it necessary to search over a range of both frequencies and frequency 
derivatives. Fully coherent blind searches for $\gamma$-ray pulsars are therefore extremely computer intensive, since the number of frequency bins 
in the FFT increases with the length of the observational time ($N_{\rm{bins}} = 2 T f_{\rm{max}}$), where $T$ is the duration of the 
observation and $f_{\rm{max}}$ is the maximum search frequency~\citep{2001ApJ...556...59C}. A fully coherent pulsar search, for example, would 
require the computation of hundreds of thousands of Gigapoint FFTs to cover the spin-down range of the majority of young $\gamma$-ray pulsars. 
Such a search would also be highly sensitive to timing irregularities (e.g. timing noise and glitches).
Instead, calculating the FFT of the arrival time differences, up to a maximum time difference of order $\sim$week, greatly reduces the number of
bins in the FFT. By doing this, we are able to reduce dramatically the computational demands (both in processor and memory), as well as the 
required number of trials to span the same parameter space, while at the same time reducing our sensitivity to timing noise. All of this is 
achieved with only a modest reduction in the sensitivity, relative to fully coherent techniques~\citep{atwood06}. For these searches, we used a 
maximum time difference of $2^{19}~\mathrm{s}$ ($\sim6$ days). We binned our difference search at a time resolution of 7.8 ms, resulting in a Nyquist 
frequency of 64 Hz for our searches and searched through 2000 steps of width $\Delta ({\dot f}/{f}) = 5 \times 10^{-15}$ s$^{-1}$ to cover roughly the 
spin-down range up to that of the Crab pulsar ($-1.0 \times 10^{-11} \,\mathrm{s^{-1}} < {\dot f}/{f} < 0 \,$s$^{-1}$).

The data analyzed in this paper were collected from sky survey mode
observations beginning on 2008 August 4 (MET\footnote{Mission Elapsed Time (MET), the number of seconds since the reference time of 
January 1, 2001, at 0h:0m:0s in the Coordinated Universal Time (UTC) system, corresponding to a Modified Julian Date (MJD) of 51910 
in the UTC system.} = 239557414, MJD = 54682) and ending on 2009 July 4 (MET = 268411953, MJD = 55016). We used events with the most stringent background 
cuts (\emph{diffuse} class photons of Pass 6.3~\citep{LATinstrument}) with zenith angle $\le 105^\circ$ (to avoid photons from the Earth's limb), and rocking
angle\footnote{The rocking angle is the angle between the pointing direction of the LAT and the zenith direction, defined as the 
direction along a line from the center of the earth through the spacecraft.} $\leq40^{\circ}$. In addition, a few 
minutes were excised around two bright gamma-ray bursts (GRBs 080916C and 090510). We applied the time-differencing technique to photons with energies 
above 300 MeV, selected from a 0.8--0.9$^\circ$ (see Table 3) circular region of 
interest (ROI) around selected target positions given by the LAT first source catalog \citep{LATCatalog}. We corrected the photon event times from each source to the solar 
system barycenter using the {\it Fermi} Science Tool \texttt{gtbary}, assuming all photons come from the target position. We searched $\sim$650 source positions from a preliminary 
version of the {\it Fermi} LAT First Source Catalog\footnote{Note that at the time we carried out our searches, the 8 sources in which we found these pulsars were not known to 
contain a pulsar and their position was therefore much less well determined.}~\citep{LATCatalog} that were not coincident with likely active galactic nuclei (AGN).

After finding a significant signal in the initial search, we followed
up on the candidate signal by performing an epoch-folding search over
a narrow region of frequency and frequency-derivative space using the
PRESTO \citep{RansomThesis} pulsar software suite.  All eight pulsars were 
discovered within $\sim11$ months of routine survey observations. They have since 
been confirmed with several months of additional data. A pulsar detection is confirmed 
if adding the new data and folding using the original timing solution results in a continued increase in the significance of the detection 
of the pulsation, as measured by the chi-squared obtained from fitting the pulse profile to a constant.

Compared to previous searches, the ones reported here are considerably more sensitive. In particular, these searches benefit from more than twice the amount of data, and a much 
better LAT source localization, than those that resulted in the discovery of the first 16 blind search $\gamma$-ray pulsars~\citep{LATBSPs}. The names and locations of the new pulsars 
are given in Table \ref{tab:names}, along with known associations, including those from the {\it Fermi} LAT First Source Catalog~\citep{LATCatalog} and the Bright 
Gamma-ray Source List~\citep{LATBSL}

\subsection{Timing Analysis}

We have generated phase-connected pulse timing models for each of the pulsars. To do this, we first extracted photons for each source using 
a radius and energy cut chosen to optimize the signal to noise ratio for each pulsar. We corrected the photon event times to the geocenter 
using \texttt{gtbary} in its geocentric mode. We then determined a set of times of arrival (TOAs) by first 
dividing our data into segments of approximately equal duration\footnote{The number of days of LAT data used to generate each TOA varied for the different pulsars, ranging from 
$\sim$9 days for PSR\,J2055+25 to $\sim$23 days for PSR\,J1846+0919. See Table~\ref{timing_solutions}.} and then folding the photon times using a preliminary ephemeris to generate 
a set of pulse profiles. The TOAs were then measured by cross-correlating each pulse profile with a kernel density template that was derived from fitting the full
mission data set, as described in \citet{TimingPaper}. Finally, we used \textsc{Tempo2}~\citep{hobbs06} to fit the TOAs to a timing model that included position, frequency, 
and frequency derivative. In the case of PSR\,J1023--5746, our timing model also included second and third derivative terms, as well as a glitch, as described below. The 
timing solutions, along with the number of days of data and TOAs that went into generating them, and the rms of the resulting model, are given in 
Table \ref{timing_solutions} (with the positions in Table \ref{tab:names}). In the case of PSR\,J1413$-$6205, there is an apparent glitch at $\sim$ MJD 54735. Periodicity searches 
of the data before the glitch indicate that the magnitude was ${\Delta f}/{f} = 1.7 \times 10^{-6}$, fairly large, though not unusually so for young Vela-like 
pulsars~\citep{dodson02,sazparkinson09}. However, the short span of data before the glitch prevented us from including the glitch in the timing 
model and thus the model for J1413--6205 in the table is constructed only from post-glitch data. 

The case of PSR\,J1023--5746 is especially complicated. As is often the case in young pulsars~\citep{hobbs04}, this pulsar suffers from large timing irregularities. 
In addition to experiencing a glitch of magnitude ${\Delta f}/{f} \sim 3.6 \times 10^{-6}$ at $\sim$ MJD 55041, this pulsar has a level of timing noise that required 
the use of higher order frequency derivative terms to model. Our model that fitted for position required terms up to $\dddot{f}$ to obtain featureless residuals to 
the fit and resulted in a position that is $\sim$2\arcsec from the {\it Chandra} X-ray source CXOU\,J102302.8--574606. To test whether the proposed association with 
CXOU\,J102302.8--574606 was compatible with the timing measurements, we made a fit with the position fixed at that of the {\it Chandra} source. This model required terms up 
to $\ddddot{f}$, but also resulted in a good fit with essentially featureless residuals. The high order polynomial terms required to model the timing noise are strongly covariant 
with the position, resulting in the statistical errors being a significant underestimate of the true error.
The \textsc{Tempo2} timing models used in this paper will be made available online at the FSSC web site\footnote{{\tt http://fermi.gsfc.nasa.gov/ssc/data/access/lat/ephems/}}.
Figure 1 shows the distribution in frequency and frequency derivative of the new pulsars (shown with unfilled triangles), compared with the previously 
known $\gamma$-ray selected pulsars (shown as solid triangles) \citep{LATBSPs} and the total known pulsar population. 

Figures \ref{1022_lightcurve}-\ref{2055_lightcurve} show the folded light curves (two rotations, for clarity, and 32 bins per rotation) of the eight new pulsars 
in five energy bands ($>$ 100\,MeV, 100--300\,MeV, 300--1000\,MeV, $>$ 1\,GeV, and $>$ 5\,GeV), ordered by increasing right ascension. 
Each band illustrates the peak structure across different energies. The phases have been shifted so that the ``first" peak lies approximately at $\phi$=0.25. 
The name of each peak (i.e. ``first" vs ``second") refers to their respective order of arrival relative to the off-peak emission. We caution, however, that for some 
pulsars (e.g. PSR\,J1954+2836) there may be no appreciable difference between the off-peak emission and the ``bridge" emission between the 
two peaks, leading to some ambiguity in such designation. Consequently, there is also some ambiguity in our measurement of the corresponding peak separation. Five of the 
eight pulsars have a clear double peak structure, while three of them (PSR\,J1846+0919, PSR\,J1957+5033, and PSR\,J2055+25) are currently dominated by a single broad peak. 
For those pulsars showing a single peak, we tried folding the events at half or a third of the frequency (in case the original frequency found in our search was the second or 
third harmonic), but the various peaks obtained in this way were not statistically distinguishable. The additional statistics accumulated from continued observations by the LAT 
are necessary to firmly establish whether the single broad peaks are real, or whether they can be resolved into narrow multiple peaks with small peak separations.

The pulse shape parameters, including peak multiplicity, $\gamma$-ray peak separation (computed by fitting the peaks with two gaussians and taking 
the difference between the means), and off-peak definition, are given in Table \ref{pulse_shape_parameter_table}.

\subsection{Spectral Analysis\label{spectral}}

We have performed the same spectral analysis on the eight pulsars as was performed on the 46 pulsars presented in the 
First LAT Catalog of $\gamma$-ray Pulsars~\citep{LATPulsars}. The pulsar spectra were fitted with an exponentially cut off power law model
of the form:

\begin{equation}
\frac{dN}{dE} = K E^{-\Gamma}_{\rm{GeV}} e^{-\frac{E}{E_{\rm{cutoff}}}}
\end{equation}

where the three parameters are the photon index $\Gamma$, the cutoff energy $E_{\rm{cutoff}}$, and a normalization factor $K$ (in units 
of ph cm$^{-2}$ MeV$^{-1}$), defined at an energy of 1 GeV. In order to extract the spectra down to 100 MeV, we must take into account 
all neighboring sources and the diffuse emission together with each pulsar. This was done using a 6-month source list that was generated 
in the same way as the Bright Source List, following the prescription described in \cite{LATPulsars}. We used all events in an ROI of 
10$^\circ$ around each pulsar, and included all sources up to 17$^\circ$ into the model (sources outside the ROI can contribute at low energy). 
The spectral parameters for sources outside a 3$^\circ$ radius of the pulsar were frozen, taken from the all-sky analysis, while those for the 
pulsar and sources within 3$^\circ$ were left free for this analysis. In general, nearby $\gamma$-ray sources are modeled by a simple power law. In the case 
of PSR\,J1023--5746, however, the nearby sources include the $\gamma$-ray pulsar PSR\,J1028--5819~\citep{LATPSR1028}, as well as the bright pulsar-like LAT source 
1FGL\,J1018.6-5856, both of which were modeled by a power law with an exponential cutoff. For the eight pulsars, an exponentially cut off power law spectral model was 
significantly better than a simple power law. This can be seen by computing the test statistic $TS_{\rm{cutoff}}$ = 2$\Delta$log(likelihood) of the model with a cutoff 
relative to one with no cutoff, as shown in column 7 of Table~\ref{spectral_parameters}, where the lowest value of $TS_{\rm{cutoff}}$ is $\sim$20.
Because the $\gamma$-ray sources we are fitting might contain a significant unpulsed component, we attempt to improve our fits to the pulsar spectrum by 
splitting the data into an on-peak and off-peak component, and performing the fit only to the on-peak events, while using the off-peak component to 
better estimate our background. Table~\ref{pulse_shape_parameter_table} gives the definition of the off-peak phase intervals (the on-peak interval being the complement of the off-peak one), 
as determined by visual inspection of the light curves in all energy bands \citep[the same method used in][]{LATPulsars}. 
We used the off-peak window to estimate the unpulsed emission. We re-fitted the on-peak emission to the exponentially cut off 
power law, with the off-peak emission (scaled to the on-peak phase interval) added to the model and fixed to the off-peak result. The resulting 
spectral index and energy cutoff for the on-peak emission of each pulsar are those given in columns 4 and 5 of Table~\ref{spectral_parameters}. For more 
details on the spectral fitting see \cite{LATPulsars}.

Three of the pulsars (J1023--5746, J1413--6205, and J2055+25) showed significant emission ($TS>$25) in the off-peak component. The 
off-peak energy spectrum of PSR J1023--5746 shows no indication of a cutoff. Furthermore, the distribution of this emission does not follow the Galactic plane (as might 
be expected if this were due to an improperly modeled diffuse background) and peaks close to the pulsar position, all of which suggests that 
such emission may be due to a pulsar wind nebula (PWN). However, the other two pulsars, J1413--6205 and J2055+25, have an off-peak component that is best fit by a power law with 
exponential cutoff, with photon index of $\Gamma=1\pm0.6$ and $\Gamma=1\pm0.5$, and $E_{\rm{cutoff}}=1.8\pm1$\,GeV and 
$E_{\rm{cutoff}}=0.65\pm0.26$\,GeV respectively. This could be an indication of magnetospheric origin, similar to what is seen in the case of 
PSR\,J1836+5925~\citep{PSRJ1836}. We caution, however, that these results are subject to some caveats. In the case of J1413--6205, the significance 
of the preference of the power law with a cutoff over a simple power law is marginal (a $TS$ of 36 vs 26). The case of J2055+25, on the other hand, is 
highly sensitive to our definition of the off-peak phase interval. A reduction of the off-peak interval from 0.66 to 0.5 (i.e. a 25\% reduction) 
reduced the $TS$ from 99 to 52 (nearly 50\%), which seems to suggest that we may be overestimating the ``true" size of the off-peak interval. 
After obtaining the spectral parameters for each pulsar, we compute the corresponding photon and energy fluxes, and present the results in 
columns 2 and 3 of Table~\ref{spectral_parameters}.
It is interesting to note that the oldest pulsars tend to have the hardest spectra and lowest cutoff energies. Earlier studies in \cite{LATPulsars} showed 
a weak trend of spectral index with $\dot E$, and a stronger trend is evident here, albeit with lower statistics.

\section{The New $\gamma$-ray Pulsars}

The new pulsars are drawn from similar populations as the first 16 pulsars discovered in blind searches. Five of the pulsars 
(J1023--5746, J1044--5737, J1413--6205, J1429--5911, and J1954+2836) are energetic ($\dot E \ge 7 \times 10^{35}$ erg s$^{-1}$) and 
young ($\tau_c < 100$\,kyr), with a large magnetic field at the light cylinder ($B_{LC} \ge 10$ kG). The remaining three 
(J1846+0919, J1957+5033, J2055+25) have smaller $B_{LC} \sim$ 0.3--1 kG, are less energetic ($\dot E \sim (0.5-3) 
\times 10^{34}$ erg s$^{-1}$) and have much larger characteristic ages ($\tau_c > 300$ kyr). 
\subsection{Source Associations}

As shown in Table~\ref{tab:names}, five of the eight $\gamma$-ray pulsars are associated with sources found in the {\it Fermi} bright source 
list~\citep{LATBSL}, while two of the pulsars were also reported as $\gamma$-ray sources by the AGILE team~\citep{AGILEcatalog}. 
Only PSR\,J1413--6205 was clearly detected as a $\gamma$-ray source by EGRET 
(EGR\,J1414--6224~\citep{EGRcatalog}\footnote{Given the complicated nature of this region, it is safe to assume that at least part of the $\gamma$-ray emission from 
PSR\,J1413--6205 was captured as 3EG\,J1410--6147~\citep{3EGcatalog}, even if PSR\,J1413--6205 falls slightly outside the EGRET 95\% statistical error 
error circle.}) although it is very likely that another one, PSR\,J1023--5746, was at least partly responsible for the EGRET source 
3EG\,J1027--5817~\citep{LATPSR1028}. Three pulsars (PSRs\,J1023--5746, J1413--6205, and J1954+2836) may have even been detected as sources by the COS-B detector 
(2CG\,284--00, 2CG\,311--01, and 2CG\,065+00 respectively~\citep{COSB2catalog}), though it is difficult to establish a one-to-one correspondence between 
such sources and the pulsars, since the COS-B error radii are quite large ($\sim1^\circ$) and source confusion becomes an issue. It appears, for example, 
that the COS-B source 2CG\,284-00 was made up of various contributions which the LAT has now resolved into three separate $\gamma$-ray sources, including 
the pulsars PSR\,J1028--5819 and the newly-discovered PSR\,J1023--5746~\citep{LATPSR1028}. We should point out that the issue of source confusion is one 
faced by the LAT too, albeit at a different sensitivity level than EGRET. The problem is particularly severe in the Galactic plane, where there is a large
component of diffuse emission. In the case of PSR\,J1023--5746, for example, this resulted in an initial LAT source location that was far 
removed ($>10\arcmin$) from the true position of the pulsar (see Figure~\ref{LAT_image}), making the early detection of pulsations from this source 
all the more challenging.

\subsection{Distance Estimates\label{distances}}

In the absence of distance information from observations in other
wavebands, for example radio dispersion measures or optical parallax
determinations, it is still possible to discern a general idea of the
distance to these blind search pulsars. The method hinges upon the
observed correlation between intrinsic $\gamma$-ray luminosity $L_{\gamma}$
above 100 MeV and pulsar spin-down energy loss rate ${\dot E}$, as
depicted in Figure 6 of \cite{LATPulsars}.  
The luminosity trend is calibrated from the observed $> 100$
MeV fluxes using standard radio pulsar distance determination techniques
and the presumption of a beam correction factor\footnote{The beam correction factor, $f_\Omega$, allows us to convert the 
observed flux $F_{obs}$ into $\gamma$-ray luminosity, $L_\gamma$, using the formula $L_\gamma$=4$\pi$$f_\Omega$$F_{obs}$$D^2$, where D is the distance.} $f_\Omega= 1$  
for the $\gamma$-ray emission cone for all pulsars. 

The correlation is benchmarked at $L_{\gamma}\sim 3.2 \times 10^{33} ({\dot E}_{34})^{1/2}$ erg s$^{-1}$ in
Figure 6 of \cite{LATPulsars}, where ${\dot E}_{34}$ is ${\dot E}$ in units of $10^{34}$ erg s$^{-1}$. Note that for $\dot{E}\lesssim 10^{34}$ erg s$^{-1}$ there is a 
break in the observed trend to a stronger dependence on $\dot{E}$ (e.g. L$_\gamma \sim \dot{E}$~\citep{LATPulsars}. This is expected as pulsars cannot exceed 
100\% radiative efficiency. For the two pulsars that are below this transitional $\dot{E}$ (PSR\,J1957+5033 and PSR\,J2055+25), the efficiencies will be overestimated and 
thus the pseudo-distances given in Table~\ref{spectral_parameters} may be overestimates.

There is considerable scatter (by factors of 3-10) of inferred pulsar $\gamma$-ray luminosities about this
linear relationship, possibly due to the inaccuracy of the
$f_\Omega= 1$ assumption. An earlier version of this correlation was
calculated for the EGRET pulsar database~\citep{thompson99}.  Physical
origins for such a relationship are discussed in \cite{zhang00}, having first been 
identified by \cite{harding81}.

For a blind search pulsar with unknown dispersion measure, we can use this correlation, along with our determination of ${\dot E}$ from 
the measured spin-down parameters, to specify $L_{\gamma}$, and then invert it to obtain a measure of its distance. This is referred to as 
the $\gamma$-ray pulsar {\it pseudo-distance} $d_{ps}$, and incorporates 
the presumption of $f_\Omega\sim 1$ in the absence of refined estimates of the solid angle, which can be obtained using pulse profile 
information~\citep{watters09}. In this way, one obtains
\begin{equation}
d_{ps} = 1.6 \frac{(\dot{E}_{34})^{1/4}}{(f_\Omega G_{100-11})^{1/2}}\,\mathrm{kpc} \\ 
\end{equation} 
Here $G_{100-11}$ is the observed energy flux $G_{100}$, as listed
in Table~\ref{spectral_parameters}, in units of $10^{-11}$ erg cm$^{-2}\,$s$^{-1}$.
The significant scatter of inferred luminosities for radio-loud LAT
pulsars about the correlation translates to uncertainties in
pseudo-distances of the order of factors of 2--3.  The resulting
$d_{ps}$ estimates for the new blind search pulsars are listed in
Table 4, ranging from $d_{ps}\sim0.4$\,kpc for PSR J2055+2539
to $d_{ps}\sim 2.4$\,kpc for the high ${\dot E}$ PSR\,J1023--5746. The relatively large pseudo-distance 
for PSR\,J1023--5746 is naturally expected due to interplay between its more-or-less typical $G_{100}$ flux and its high ${\dot E}$. It is of considerable
interest because of the possible association of this pulsar with the Westerlund 2 cluster, as discussed in Sections \ref{xray} and \ref{tev}.

\section{Multiwavelength Observations}

\subsection{Radio}

These pulsars were all discovered in $\gamma$-ray searches and thus are $\gamma$-ray selected pulsars, but targeted radio observations 
are required to determine if they are also radio quiet, or could have been discovered in radio surveys independently.  The population 
statistics of radio quiet vs. radio loud $\gamma$-ray pulsars have important implications for $\gamma$-ray emission models~\citep[e.g.][]{gonthier04}.
The precise positions derived from the LAT timing of these pulsars allowed us to perform deep follow up radio observations to search 
for pulsations from each of the new pulsars.  We used the NRAO 100-m Green Bank Telescope (GBT), the Arecibo 305-m radio telescope, and the 
Parkes 64-m radio telescope for these observations.  The log of observations is shown in Table \ref{tab:radiolims} and the instrument 
parameters used in the sensitivity calculations are shown in Table \ref{tab:radioobs}.  All observations were taken in search mode and the 
data were reduced using standard pulsar analysis software, such as PRESTO~\citep{RansomThesis}. We searched for pulsations for a range of values of 
dispersion measure (DM), from 0 to some maximum, given in Table \ref{tab:radiolims}.

To compute the minimum pulsar flux that could have been detected in these observations, we use the modified radiometer equation 
\citep[e.g.][]{lorimerkramer04}:
\begin{equation}
S_\mathrm{min} = \beta \frac{(S/N)_\mathrm{min} T_\mathrm{sys}}{G \sqrt{n_\mathrm{p} t_\mathrm{int} \Delta f}} \sqrt{\frac{W}{P-W}}
\end{equation}
where $\beta$ is the instrument-dependent factor due to digitization and other effects; $(S/N)_\mathrm{min} = 5$ is the threshold signal 
to noise for a pulsar to have been confidently detected; $T_\mathrm{sys} = T_\mathrm{rec} + T_\mathrm{sky}$; $G$ is the telescope gain; 
$n_\mathrm{p}$ is the number of polarizations used (2 in all cases); $t_\mathrm{int}$ is the integration time; $\Delta f$ is the observation 
bandwidth; $P$ is the pulsar period; and $W$ is the pulse width (for uniformity, we assume $W=0.1P$).

We use a simple approximation of a telescope beam response to adjust the flux sensitivity in cases where the pointing direction was 
offset from the true direction to the pulsar.  This factor is
\begin{equation}
f = \exp\left( \frac{-(\theta/\mathrm{HWHM})^2}{1.5} \right),
\end{equation}
where $\theta$ is the offset from the beam center and HWHM is the beam half-width at half maximum.  A computed flux limit of $S$ at the 
beam center is thus corrected to $S/f$ for target offset from the pointing direction.  The resultant flux limits are compiled in Table 
\ref{tab:radiolims}. We compare these flux limits with the measured fluxes of the population of pulsars in the ATNF pulsar catalog~\citep{ATNFcatalog} 
in Figure \ref{fig:radiolims}. To make the fluxes comparable, we have scaled them all to the equivalent 1400 MHz 
flux density using a typical pulsar spectral index of 1.6. Using the scaled flux densities and the pseudo-distances given in Table~\ref{spectral_parameters}, we 
can estimate the luminosity limits at 1.4\,GHz. Seven out of the eight pulsars have luminosity limits below $L_{1.4} = 0.2$\,mJy\,kpc$^2$. The remaining one, 
PSR~J1023--5746, has a limit of $\sim0.6$\,mJy\,kpc$^2$. Prior to the detection of radio pulsations from three LAT blind search pulsars \citep{camilo09,MGRO}, 
the least luminous young radio pulsar known was PSR~J0205+6449 in SNR~3C~58 with $L_{1.4} \approx 0.5$\,mJy\,kpc$^2$ \citep{camilo02}. Two of the new radio LAT 
pulsars, however, have luminosities an order of magnitude below this level.  Thus, while the radio limits we present here are quite stringent by comparison with 
the overall known population of young pulsars, they are still far above the now least luminous pulsars known, and it is entirely possible that some of the new 
pulsars could be radio emitters at a level below our limits. It should also be noted that some of these pulsars, and in particular PSRs\,J1846+0919, J1957+5033, 
and J2055+25 located at high Galactic latitudes ($|b|>3^\circ$), may be very nearby and have very low values of DM. As 
was seen in the case of PSR\,J1741--2054~\citep{camilo09}, the received flux from such a nearby pulsar can vary greatly due to interstellar scintillation. The best way to detect 
such pulsars in radio might involve multiple low frequency observations. However, of the three pulsars mentioned, only PSR\,J2055+25 has at present been observed more than 
once (see Table~\ref{tab:radiolims}).

The lack of even a single radio detection among these eight pulsars brings the total number of known radio quiet pulsars to 22 (including Geminga), 
out of a total of 46 known young $\gamma$-ray pulsars, or $\sim$50\%. Such a high fraction may indicate that the size of the $\gamma$-ray 
beam is significantly larger than that of the radio beam, consistent with the predicitons of fan-beam outer-magnetosphere models for $\gamma$-rays 
and narrow polar-cap models for radio beams. However, more detailed population studies will be required to quantify this effect.

\subsection{X-ray\label{xray}}

We searched archival X-ray observations for possible counterparts of the newly-discovered LAT pulsars, but found the pre-existing 
X-ray coverage of these fields to be extremely sparse. PSR\,J1023--5746 is the only pulsar with any significant X-ray observations, 
due to its proximity to the Westerlund 2 cluster. It was observed on three different occasions\footnote{Obsids 3501, 6410, and 6411.} (August 2003 and September 2006) 
by \textit{Chandra} in the ACIS-S \texttt{vfaint} mode, with a total exposure of $\sim$130 ks. In a new analysis of the {\it Chandra} data, we found a faint source at 
$\alpha_{2000}$ = 10$^\mathrm{h}$ 23$^\mathrm{m}$ 02\fs8, $\delta_{2000}$ = --57\degr 46\arcmin 07\farcs01 with a 95\% error radius of 
$\sim$2\arcsec, consistent with our best timing position for the pulsar obtained from LAT data (see Figure~\ref{Chandra_image}). This source 
is one of 468 X-ray sources previously reported by \cite{tsujimoto07} in their survey of the massive star-forming region RCW 49. 
While \cite{tsujimoto07} identify it as CXOU J102302.84--574606.9, it is also referred to in the SIMBAD\footnote{{\tt http://simbad.u-strasbg.fr/simbad/}} astronomical 
database as CXOU J102302.8--574606, and we hereon after adopt this designation.

We find no optical source within 5\arcsec\, (upper limit m$_V \sim$ 21) in the NOMAD optical catalog. CXOU\,J102302.8--574606 has a typical PWN-like power law 
spectrum with a photon index of 1.2$\pm$0.2 and an $N_H$ of (1.5$\pm0.4)\times10^{22}$ cm$^{-2}$, resulting in an 
unabsorbed flux of 1.2$\pm0.3\times10^{-13}$ erg cm$^{-2}$ s$^{-1}$ (for a total of $\sim$600 counts). No other simple 
spectral model was statistically suitable. The off-axis PSF of the {\it Chandra} ACIS camera is a complicated function of energy, and given that 
this source is $\sim8'$ from the center of the field of view we cannot make any claims about the spatial extension of the counterpart. 
The N${_H}$ value is comparable to the Galactic one ($\sim1.3\times10^{22}$ cm$^{-2}$) implying a fairly large distance 
(possibly greater than 10 kpc). The distance to the Westerlund 2 cluster has been a matter of debate, with estimates as low as 2.8 kpc~\citep{ascenso07}, or as high as 
 8.0 $\pm$ 1.4 kpc~\citep{rauw07}. The most recent estimate, based on an analysis of the CO emission and 21 cm absorption along the line of sight to the cluster places it 
at a distance of 6.0 $\pm$ 1.0 kpc~\citep{dame07}. We note that the pseudo-distance that we have estimated for PSR\,J1023--5746 (given in Table 4) is 2.4 kpc. A rough 
estimate of the average transverse velocity of the pulsar from its birth site to its current location over its lifetime is given by 
$\overline{v}_\bot \approx [(\theta_8)(d_{2.4})/(\tau_{4.6})] 1.2\times10^3$ km s$^{-1}$, where $\theta_8$ is the angular separation between the pulsar and its birth site, in 
units of 8 arcminutes, $d_{2.4}$ is the distance to the pulsar, in units of 2.4 kpc, and $\tau_{4.6}$ is the age of the pulsar, in units of 4.6 kyr. This would be among the 
highest transverse velocities measured for any pulsar~\citep[e.g.][]{hobbs04}. Accordingly, the association seems improbable unless a) the pulsar was born far from the cluster core 
(e.g. from a runaway progenitor), b) the cluster and pulsar distance are lower than our pseudo-distance estimate of 2.4 kpc, 
or c) the true pulsar age is substantially greater than its characteristic age. An on-axis {\it Chandra} observation might be able to resolve the expected bow shock structure of 
the apparent PWN and test this association.

Shortly after identifying these 8 LAT sources as pulsars, we obtained short ($\sim$5 ksec) \textit{Swift}-XRT observations in PC-mode of 5 of them 
(J1044--5737, J1413--6205, J1846+0919, J1957+5033, and J2055+25). In the first four cases, no likely X-ray counterparts were found within the 
LAT error circle. In the case of J2055+25, a 6.3\,ks observation revealed two sources relatively close to the pulsar location ($\sim$1\arcmin\,away), but 
inconsistent with the current best position derived from the timing and likely associated with bright field stars (sources b) and d) in Figure~\ref{XMM_image}). 
The upper limit on the flux for a putative X-ray counterpart in any of these five {\it Swift} XRT observations is 1.5--2$\times10^{-13}$ erg cm$^{-2}$ s$^{-1}$, under 
the hypothesis of a power law spectrum with a photon index of 2 and $N_H$ of 10$^{21}$ cm$^{-2}$. Neither PSR\,J1429--5911 nor PSR\,J1954+2836 have significant X-ray coverage.

On October 26, 2009 the {\it XMM-Newton} satellite observed the field of PSR\,J2055+25 using the three European Photon Imaging Cameras (EPIC) MOS1, MOS2, and pn instruments. For 
MOS1 and pn cameras thin filters were used while for MOS2 the medium filter was used. All three instruments obtained data in Full Frame mode which resulted in a time 
resolution of 2.6 s for the MOS1, MOS2 cameras and 73.4 ms for the pn camera.  The data from each instrument were analyzed utilizing the {\it XMM-Newton} Science Analysis System 
software (SAS) version 9.0.0. The calibration data utilized by SAS was the latest available at the time of data reduction. We filtered the event data for bad events, 
retaining only those events from the MOS cameras with predefined patterns 0-12, and also excluded times of high background. 

In Figure \ref{XMM_image} we show the combined image from the MOS1, MOS2, and pn cameras on {\it XMM}, resulting from an effective exposure of 19.0 ks after filtering. XMMU\,J205550.8+254048 and 
XMMU\,J205547.2+253906 (the ``tadpole") are apparent in the Swift XRT image but owing to the higher effective area of {\it XMM} a number of fainter sources were also detected in the image 
including a faint source virtually coincident with the pulsar position: XMMU\,J205549.4+253959. The region is crowded so in order to minimize contamination from other X-ray sources we 
form a spectrum by extracting events in a 29 arcsec circular region around the X-ray position centroid and also extract a background spectrum from a region that appears to be free of X-ray 
sources on the same CCD. For a circular region 29 arcsec in radius around a point source the MOS1 PSF encircles roughly 85\% of the photons at 1.5 keV. We can extract 
97 photons from the source region and thus can form only an approximate spectrum for XMMU\,J205549.4+253959.  Fitting an absorbed power law to this spectrum we obtain an 
approximate flux of $\sim2.4 \times 10^{-14}$ ergs cm$^{-2}$ s$^{-1}$ in the 0.5--10 keV energy band with a power law index of $\Gamma \sim 2.1$.

\subsection{TeV\label{tev}}

Four of the eight new pulsars are in the Northern Hemisphere and four are in the Southern Hemisphere. Only the Milagro TeV observatory, an all-sky 
survey instrument, has observed the four northern hemisphere pulsars. Milagro has reported a 4.3$\sigma$ excess from the location of PSR\,J1954+2836, with a measured flux at 
35 TeV of 37.1$\pm8.6 \times 10^{-17}$ TeV$^{-1}$ s$^{-1}$ cm$^{-2}$~\citep{Milagro09}. The remaining three pulsars in the Milagro field of view show no sign of TeV emission, though 
this is perhaps not surprising given that they have spin-down luminosities that are 30--200 times lower than PSR\,J1954+2836. 

Of the Southern Hemisphere pulsars, three have no reported TeV observations. The remaining one, PSR\,J1023--5746 is coincident with 
the TeV source HESS\,J1023--575 (see Figure~\ref{LAT_image}). The HESS collaboration reported the detection of TeV emission from a source with an extension 
of $\sigma = 0.18^\circ \pm 0.02$ in the vicinity of the young stellar cluster Westerlund 2~\citep{HESSJ1023}. Three possible scenarios for the TeV emission were presented by the 
HESS team: the massive WR binary system WR\,20a, the young stellar cluster Westerlund 2, and cosmic rays accelerated in bubbles or at their termination shock and interacting with their 
environment~\citep{HESSJ1023}. The estimated luminosity above 380 GeV (for an assumed distance of 8 kpc) is 1.5$\times10^{35}$ erg s$^{-1}$ or less than 1.5\% of the spin-down luminosity 
of this pulsar, making the energetics of such an association plausible \citep[e.g.][]{camilo09}. Recently, 
a jet and arc of molecular gas toward this source has been reported which suggests the possible occurrence of an anisotropic supernova 
explosion~\citep{2009PASJ...61L..23F}. The discovery of the very young and energetic pulsar PSR\,J1023--5746 suggests that it likely plays an important role in the TeV 
emission from HESS\,J1023--575. Given the long list of known TeV PWNe (currently the most numerous class of 
TeV sources\footnote{http://tevcat.uchicago.edu/}) and the significant number of these associated with bright $\gamma$-ray pulsars, such a scenario seems at least plausible, if not probable.

\section{Conclusion}
\begin{figure}
\centering
\includegraphics[width=5.5in,angle=0]{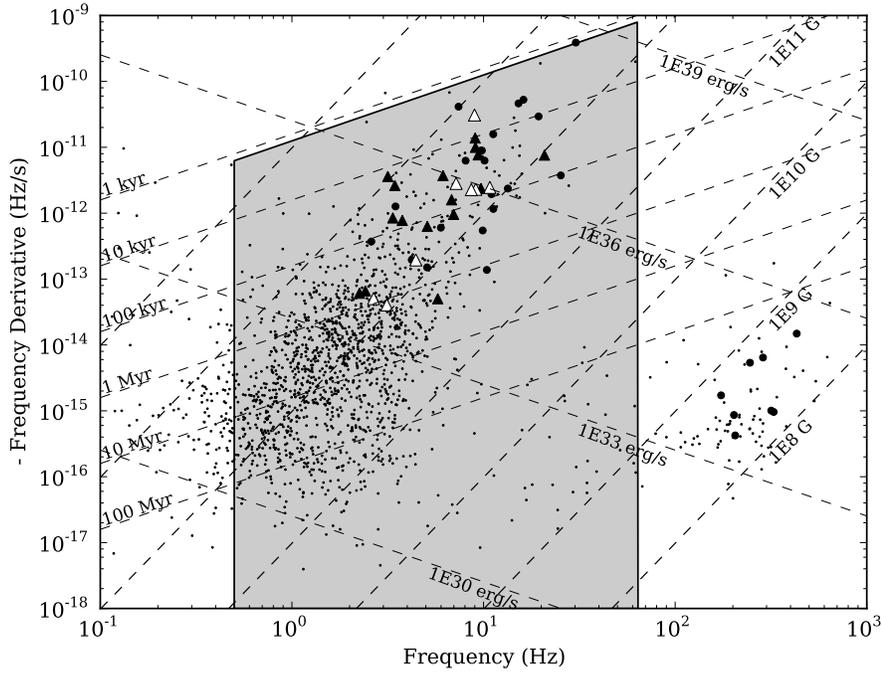}
\label{parameter_space}
\caption{The $f$--${\dot f}$ parameter space relevant to our blind searches. The pulsars in the ATNF
catalog~\citep{ATNFcatalog} are indicated by small black dots, and the radio-selected $\gamma$-ray
pulsars are indicated by larger black circles~\citep{LATPulsars}. The $\gamma$-ray selected pulsars
are indicated by triangles, where the filled-in ones denote previously reported blind search pulsars~\citep{LATBSPs}, and the unfilled
triangles denote the new pulsars reported in this paper. The shaded region corresponds to the part of the parameter space we 
have searched (see Section~\ref{bfs}). The various dashed lines indicate how the different quantities derived from the timing parameters of the pulsars 
 (characteristic age, magnetic field strength at the neutron star surface, and spin-down luminosity) vary across the parameter space. }
\end{figure}

\begin{figure}
\includegraphics[width=6in]{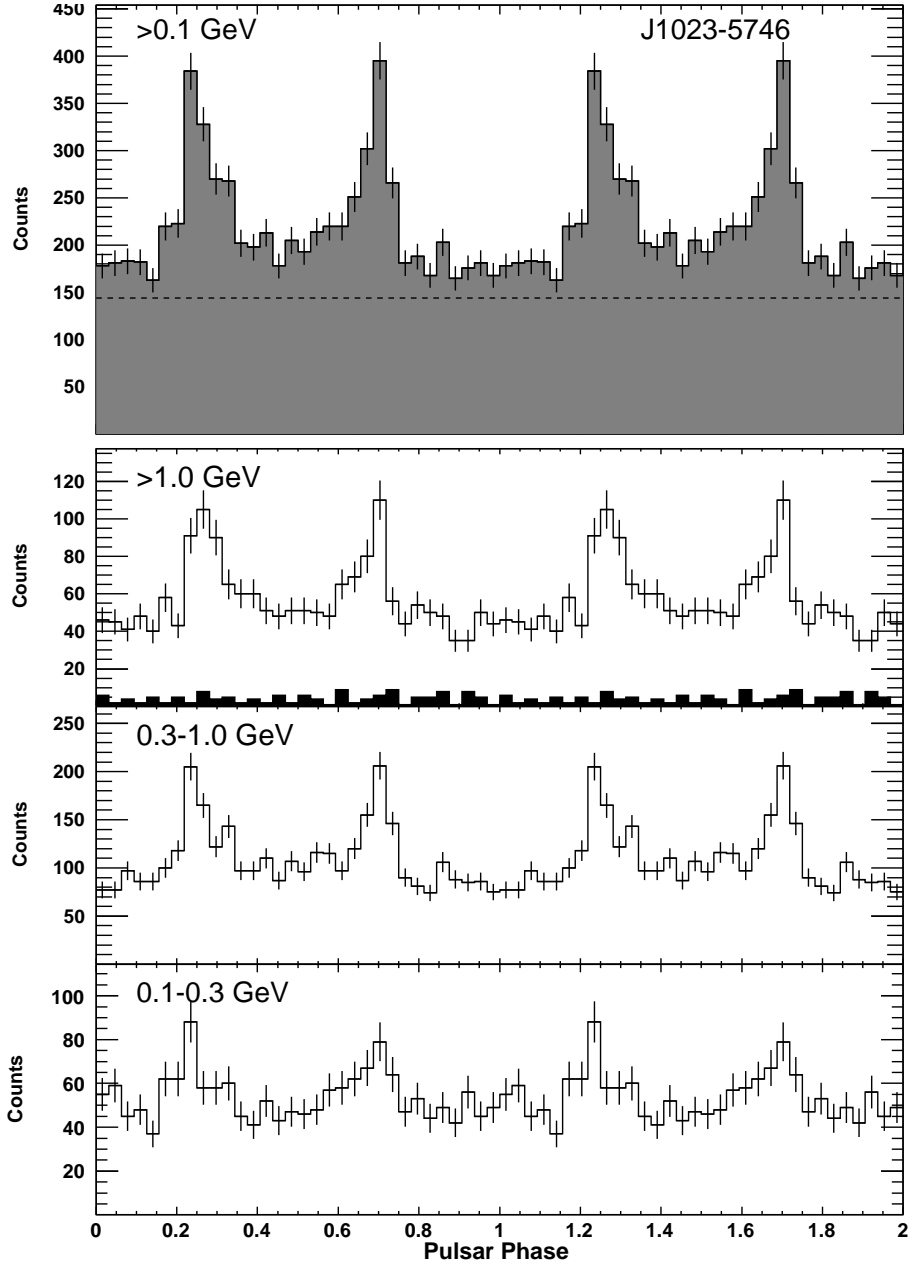}
\caption{Light curves for PSR\,J1023--5746. In this and the following figures, the folded light curves have a resolution of 32 phase bins per period. 
The black histogram in the second panel from the top shows the light curve for $E>5$ GeV. The dashed line in the top panel represents the estimated background level, as 
derived from the model used in the spectral fitting. Two rotations are shown, for clarity, and the phase of the first peak has been placed at $\sim$0.25 for clarity.\label{1022_lightcurve}}
\end{figure}

\begin{figure}
\includegraphics[width=6in]{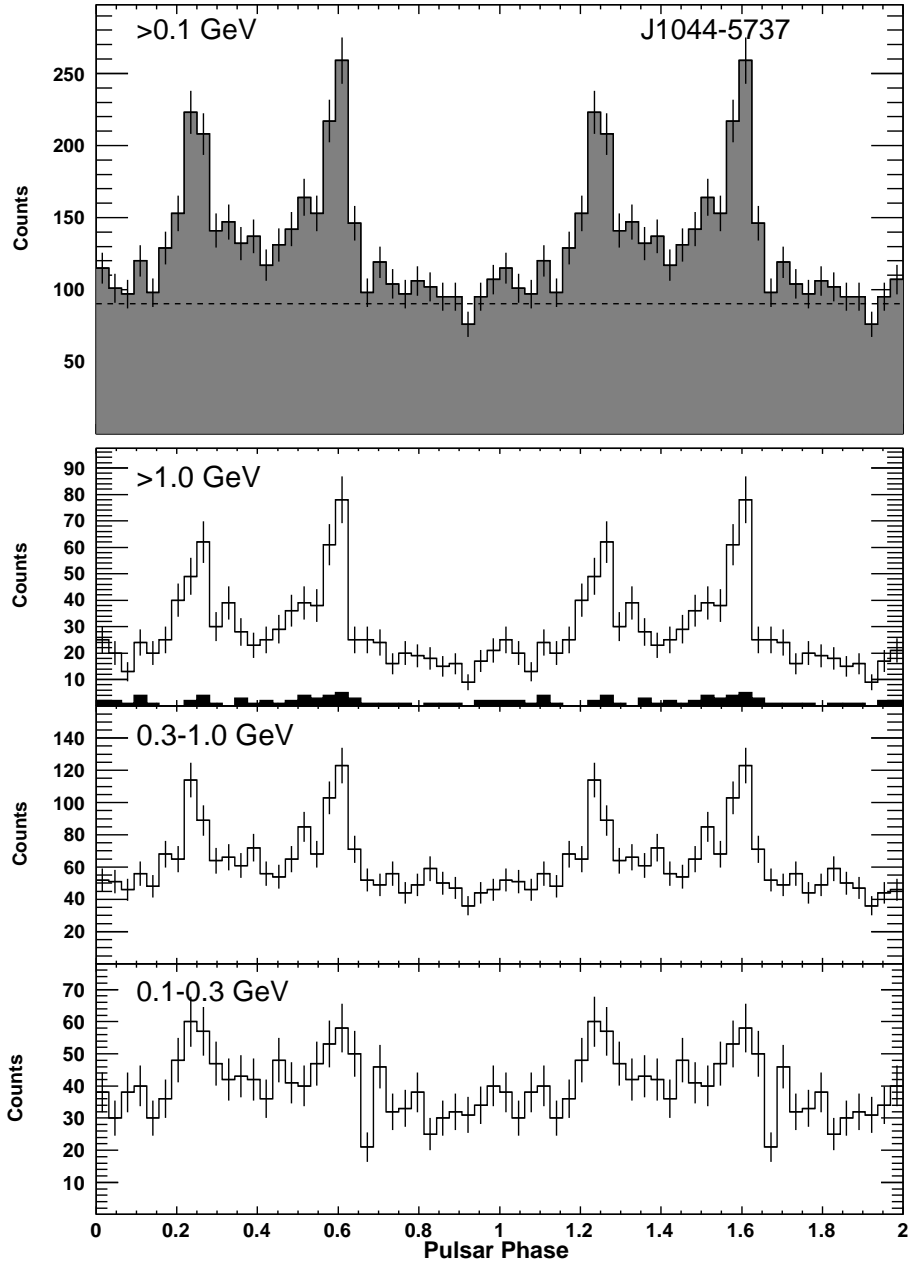}
\caption{Light curves for PSR\,J1044--5737.
\label{1044_lightcurve}}
\end{figure}

\begin{figure}
\includegraphics[width=6in]{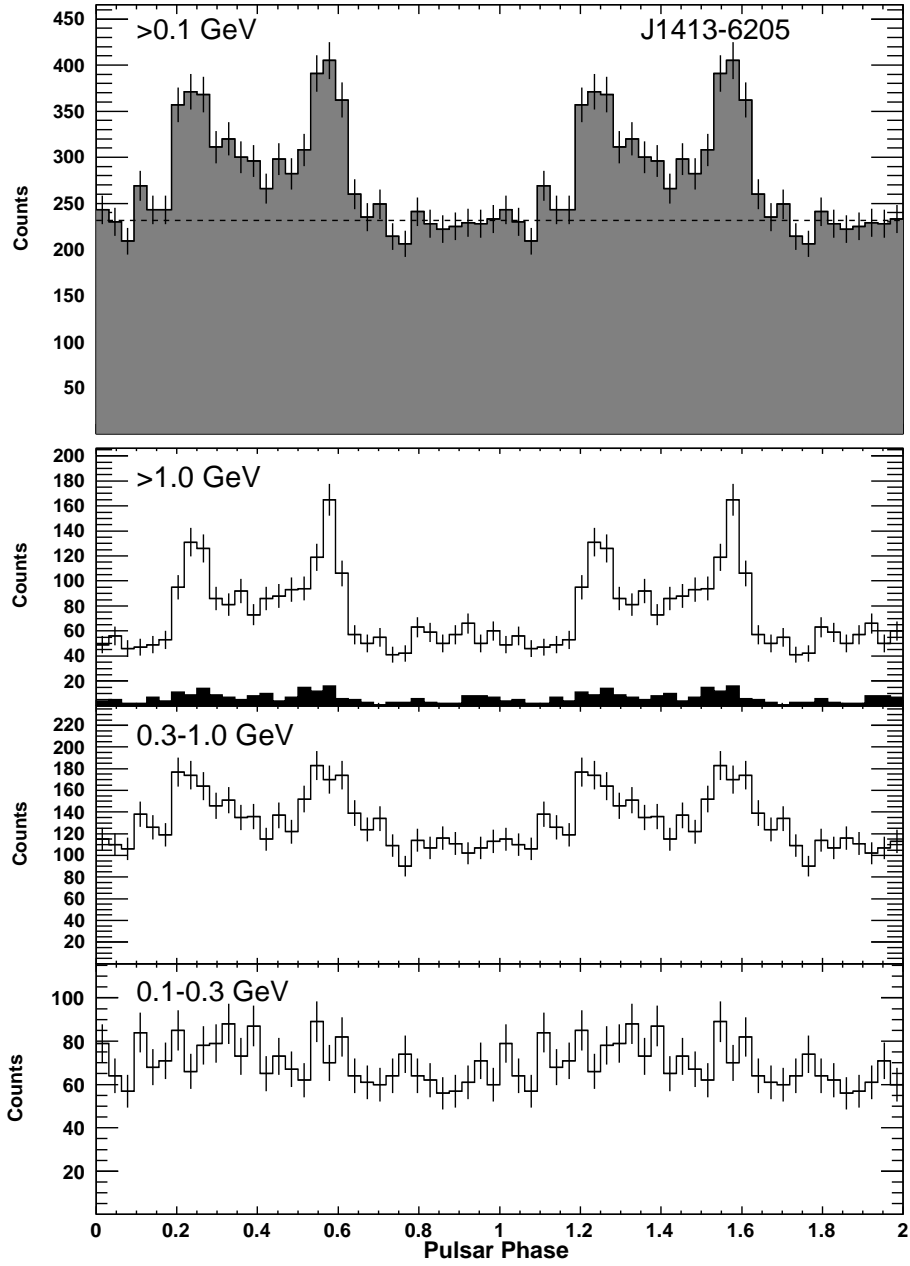}
\caption{Light curves for PSR\,J1413--6205. 
\label{1413_lightcurve}}
\end{figure}

\begin{figure}
\includegraphics[width=6in]{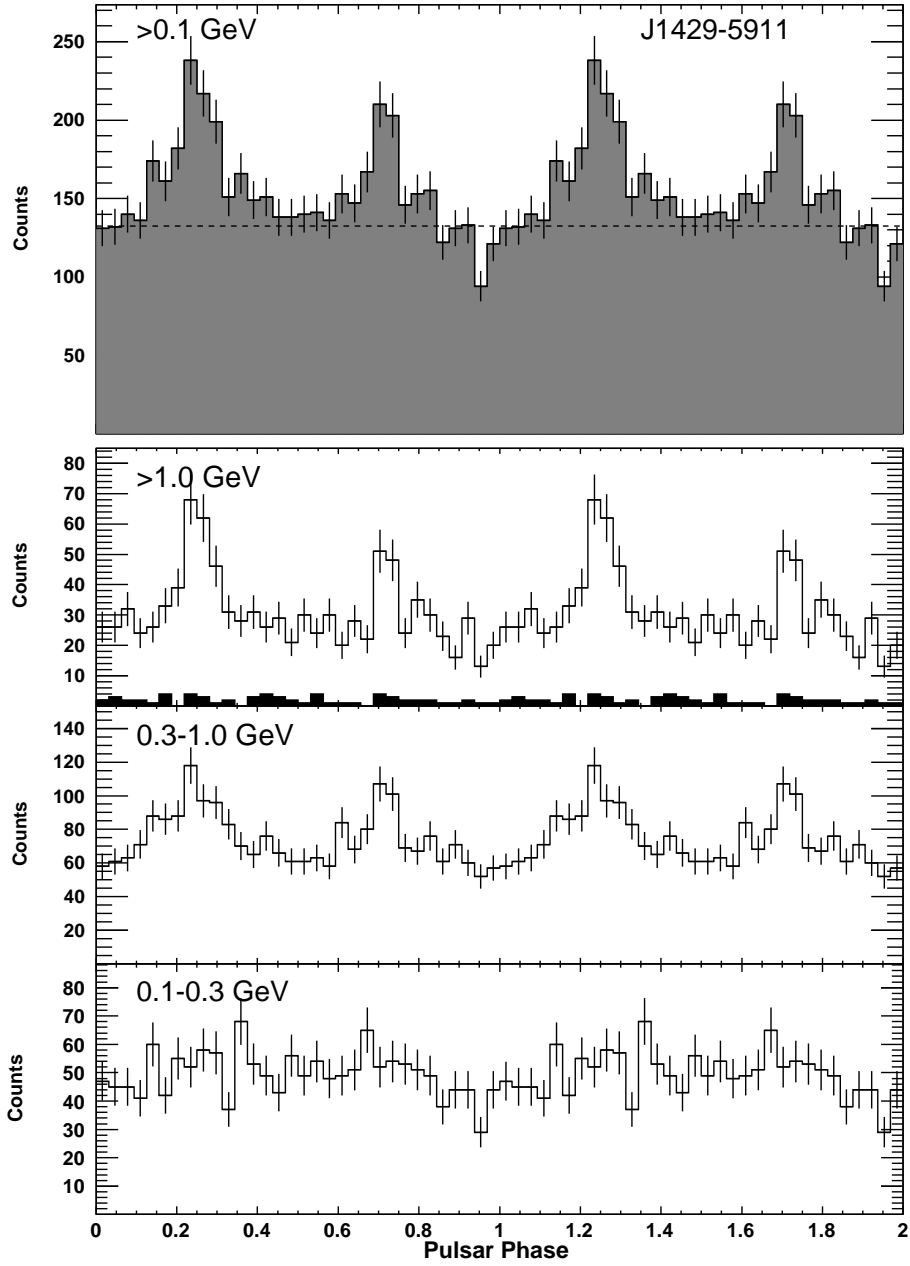}
\caption{Light curves for PSR\,J1429--5911.
\label{1429_lightcurve}}
\end{figure}

\begin{figure}
\includegraphics[width=6in]{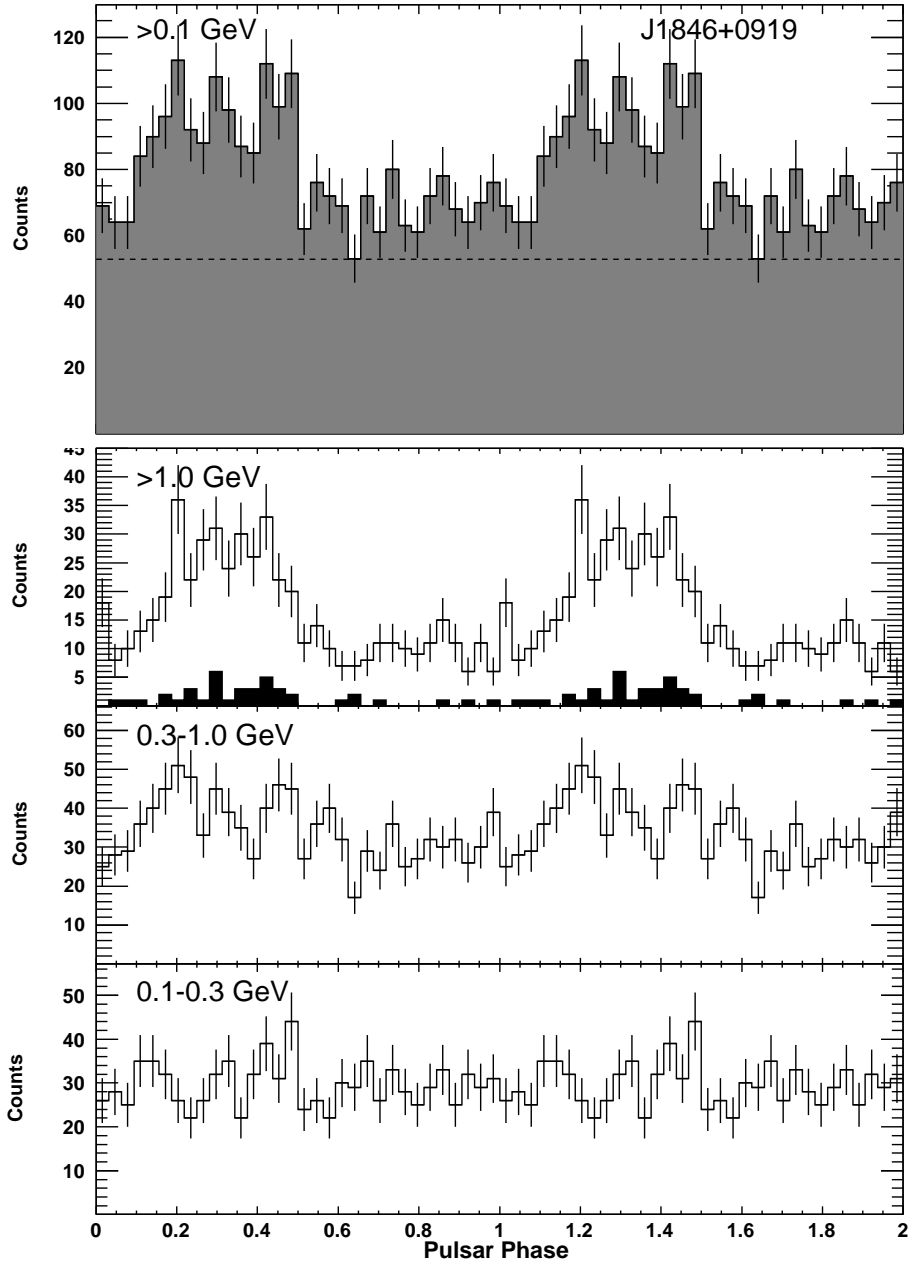}
\caption{Light curves for PSR\,J1846+0919. 
\label{1846_lightcurve}}
\end{figure}

\begin{figure}
\includegraphics[width=6in]{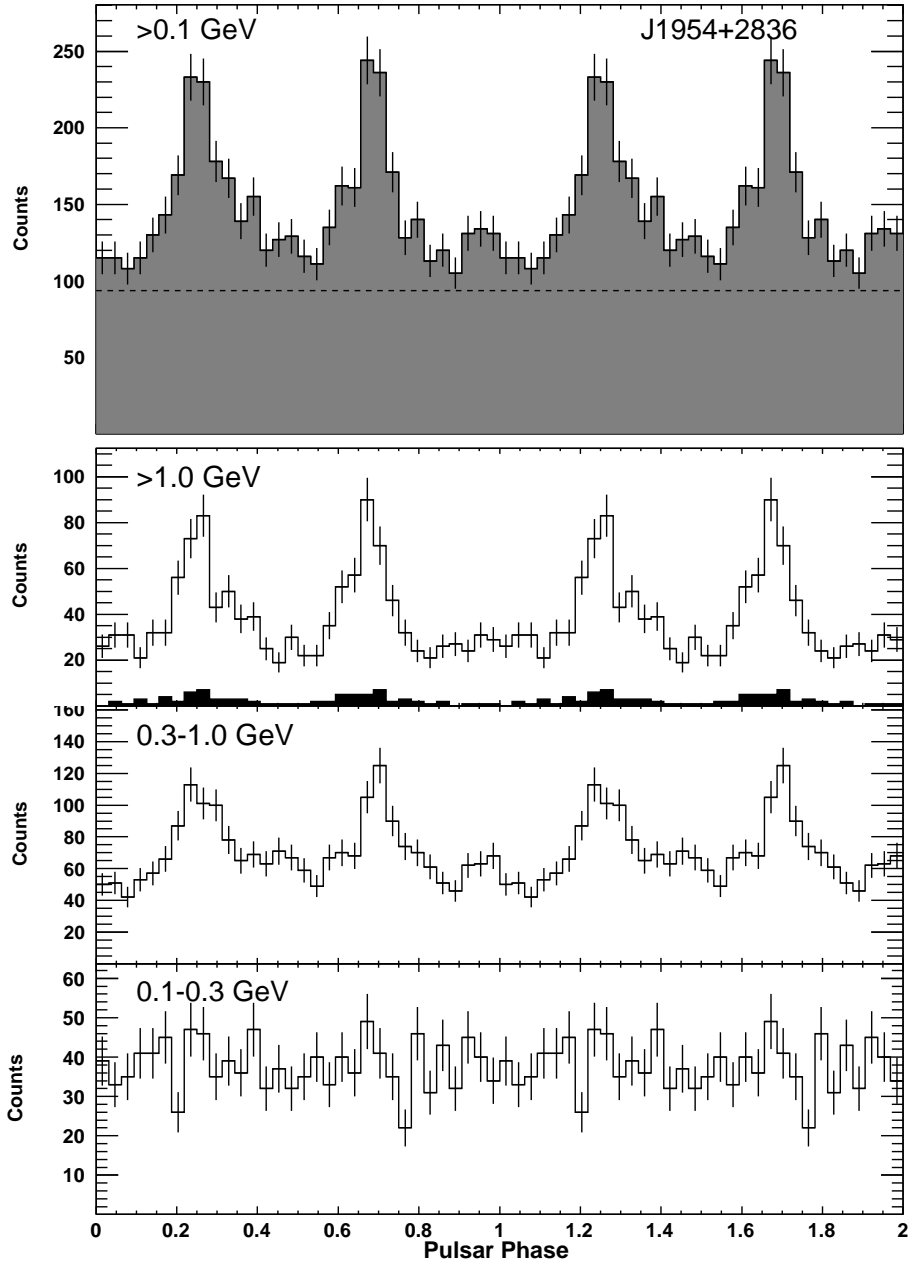}
\caption{Light curves for PSR\,J1954+2836.
\label{1954_lightcurve}}
\end{figure}

\begin{figure}
\includegraphics[width=6in]{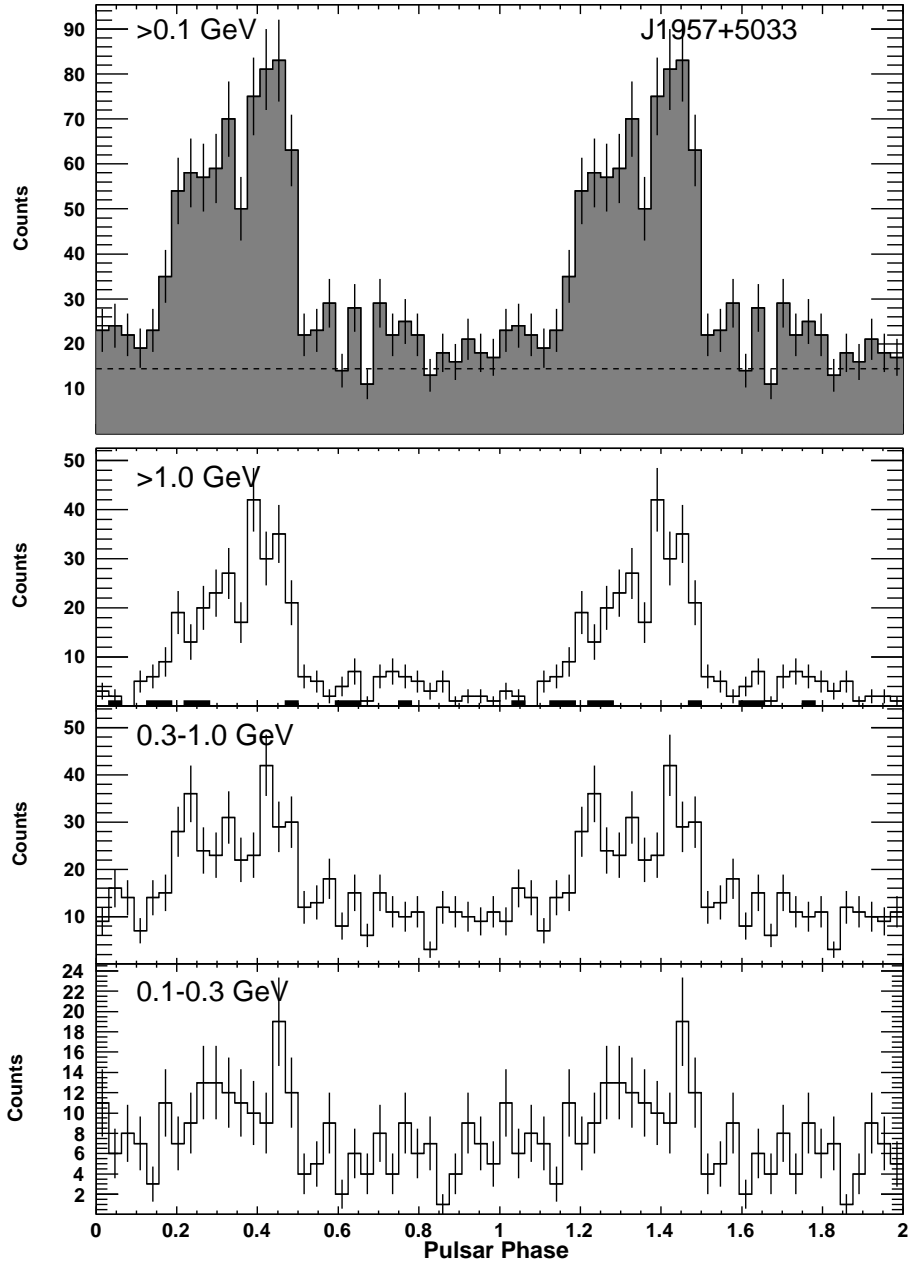}
\caption{Light curves for PSR\,J1957+5033. 
\label{1957_lightcurve}}
\end{figure}

\begin{figure}
\includegraphics[width=6in]{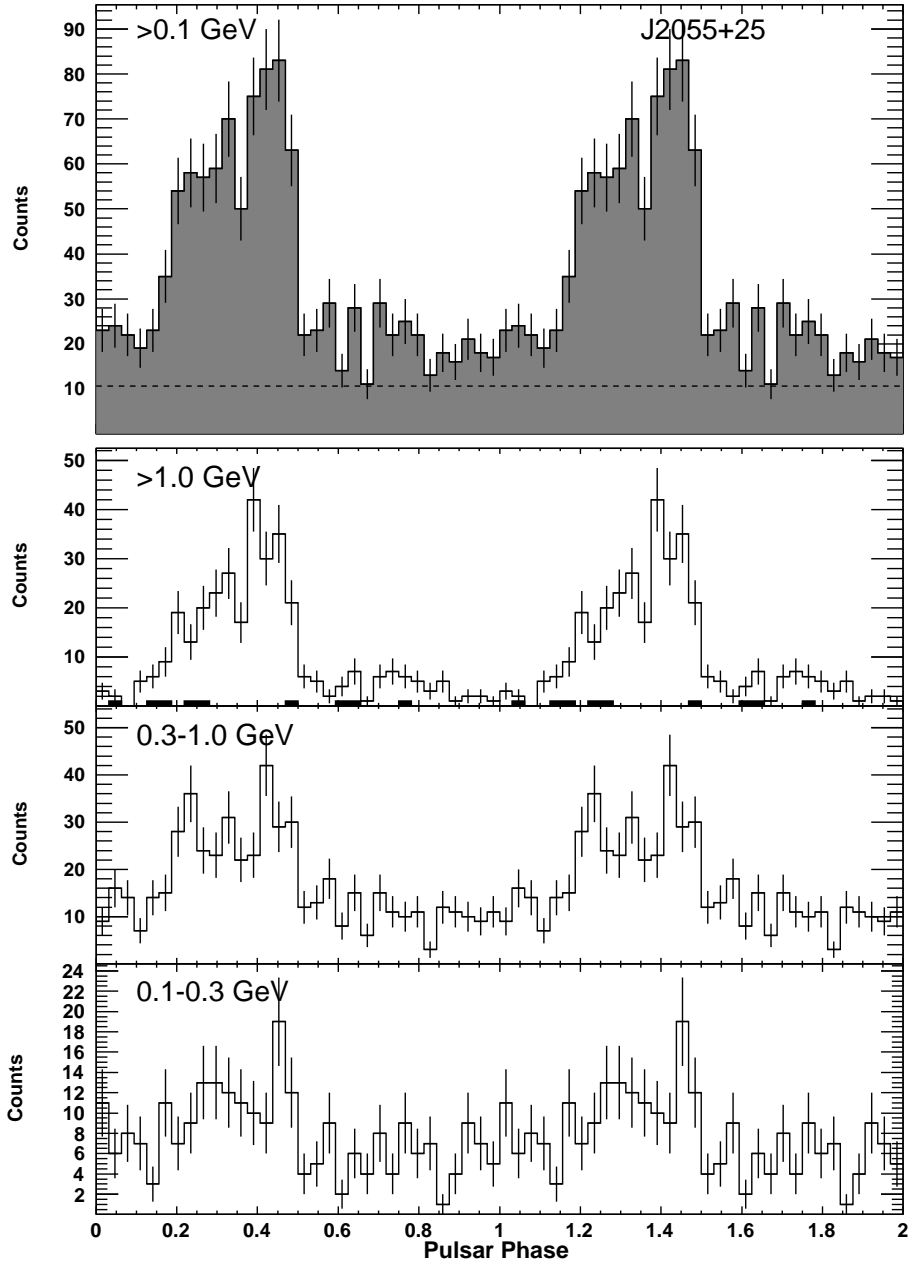}
\caption{Light curves for PSR\,J2055+25.
\label{2055_lightcurve}}
\end{figure}

\begin{figure}
\centering
\includegraphics[width=6in,angle=0]{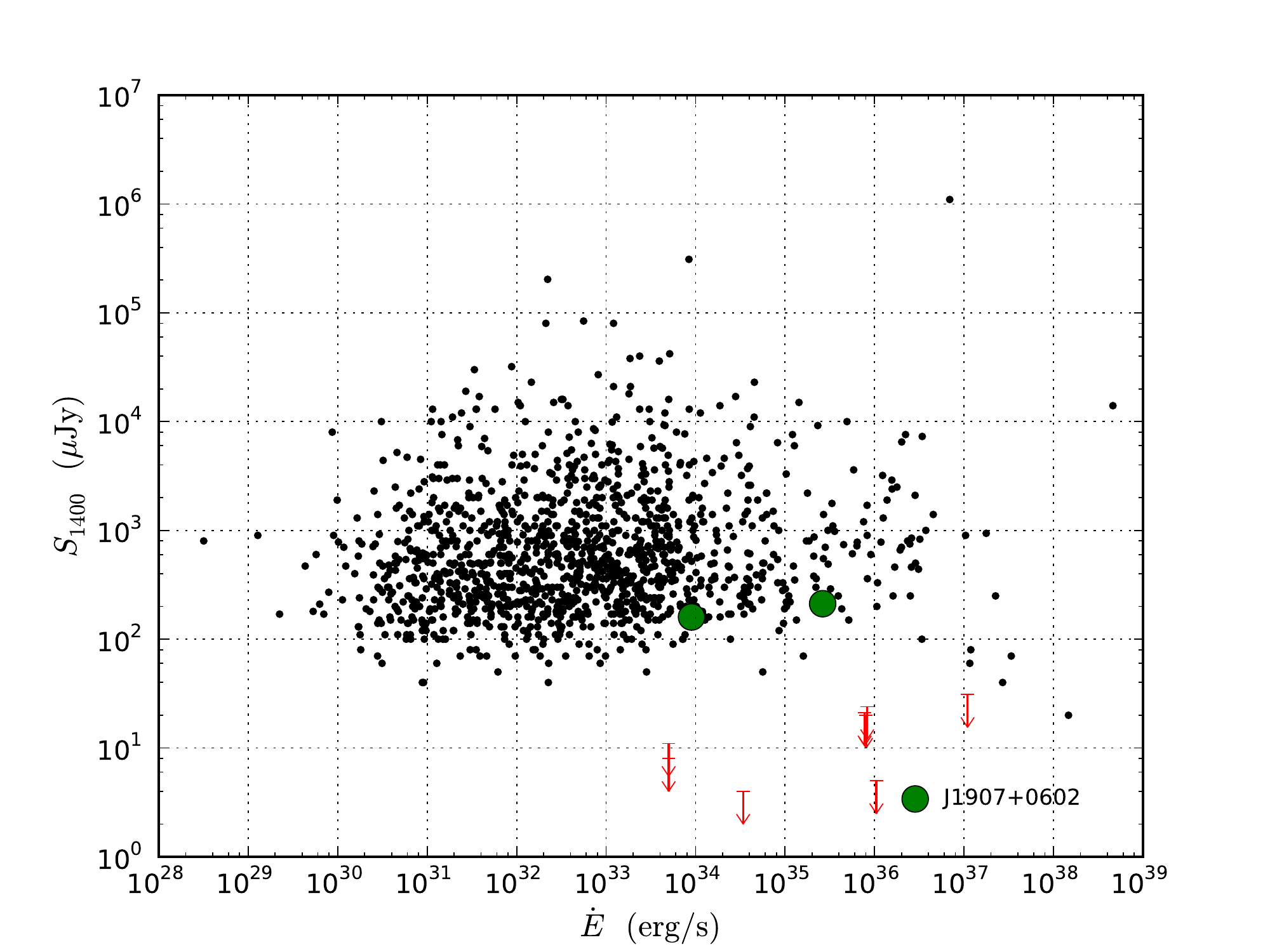}
\caption{Radio search upper limits (arrows), scaled to 1400 MHz using an assumed spectral index of 1.6, compared with the 1400 MHz flux densities from pulsars in the ATNF 
catalog~\citep{ATNFcatalog}, shown as black dots. The three larger green circles represent the radio-detected LAT blind search pulsars, including 
PSRs\,J1741--2054 and J2032+4127~\citep{camilo09}, as well as J1907+0602, the pulsar with the lowest measured flux density, first detected in a 55 minute observation 
using the Arecibo 305 m radio telescope~\citep{MGRO}.
\label{fig:radiolims}
}
\end{figure}

\begin{figure}
\epsscale{1.0}
\includegraphics[width=7in,angle=0]{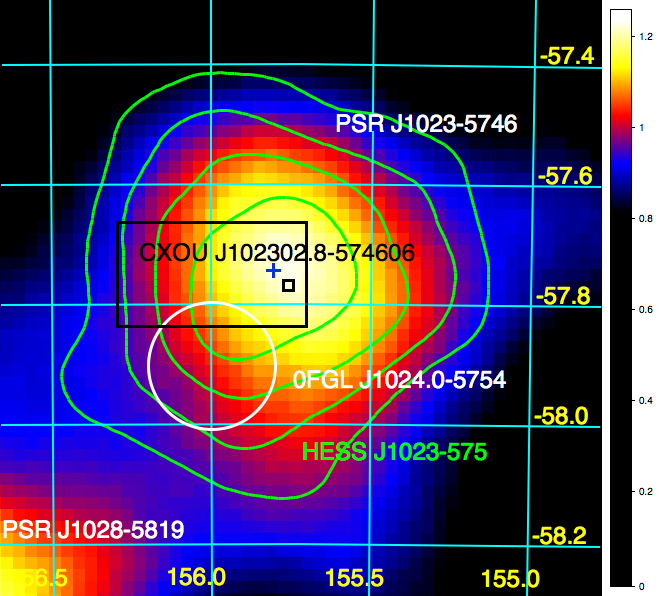}
\caption{{\it Fermi} LAT image of a 1 square degree region of the sky around PSR\,J1023--5746. The smoothed counts map was generated 
using all {\it diffuse} class events above 100 MeV between 4 August 2008 and 4 July 2009 ($\sim21.2$ Msec live time). The color 
scale is in counts per square arcminute. The overlaid green contours correspond to the significance (5$\sigma$, 6.25$\sigma$, 
7.5$\sigma$, and 8.75$\sigma$) reported by the HESS collaboration from the extended TeV source HESS\,J1023--575~\citep{HESSJ1023}. 
The LAT data are consistent with a point source. The white circle represents the position (and 95\% error circle) of 
0FGL\,J1024.0--5754, the $\gamma$-ray source reported in the {\it Fermi} Bright Source List~\citep{LATBSL}. Note that the 0FGL list 
was generated using only three months of data. The blue cross (at R.A.=155.806$^\circ$, Dec.=-57.743$^\circ$, or $\sim$2.1' away 
from the pulsar position) represents the preliminary estimate of the LAT source location based on 11 months of data and was the 
position used in the blind search that resulted in the discovery of pulsations. The large black rectangle is the area 
explored in the {\it Chandra} X-ray observations shown in Figure \ref{Chandra_image}, and the smaller black box within it, 
represents the 1 square arcminute region around the X-ray counterpart of PSR\,J1023--5746, CXOU\,J102302.8--574606, also shown 
in Figure \ref{Chandra_image} at a larger scale.
 \label{LAT_image}}  
\end{figure}

\begin{figure}
\epsscale{1.0}
\includegraphics[width=7in,angle=0]{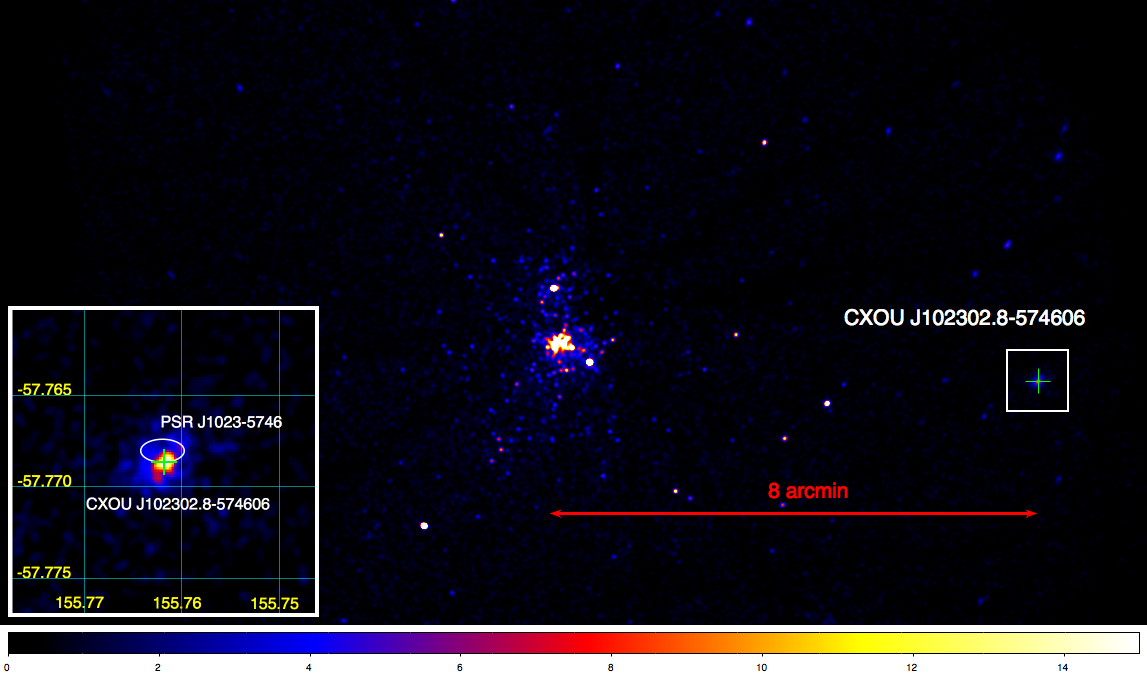}
\caption{{\it Chandra} ACIS-S (0.1--10 keV) X-ray image of the Westerlund 2 cluster region using $\sim$130 ks of data taken in August 2003 and September 2006. The color scale is in counts 
per square arcsecond. The small box on the right represents a 1 square arcminute region around the source CXOU\,J102302.8--574606, which we have identified as the X-ray counterpart of 
PSR\,J1023--5746. Note that the source is $\sim8$\arcmin\, away from the core of the cluster. {\bf Inset} -- Zoomed-in image of the 1 square arcminute region around CXOU\,J102302.8--574606. 
The white ellipse represents the 95\% confidence error ellipse of the position of PSR\,J1023--5746, based on the timing, as listed in Table~\ref{tab:names}. The errors are statistical only. 
Although the X-ray source appears to be extended, indicating the possible presence of a PWN, a full extended source analysis is dependent on a more complete understanding of the PSF of the 
instrument at such a large off-axis location.\label{Chandra_image}}  
\end{figure}

\begin{figure}
\centering
\includegraphics[width=7.5in,angle=0]{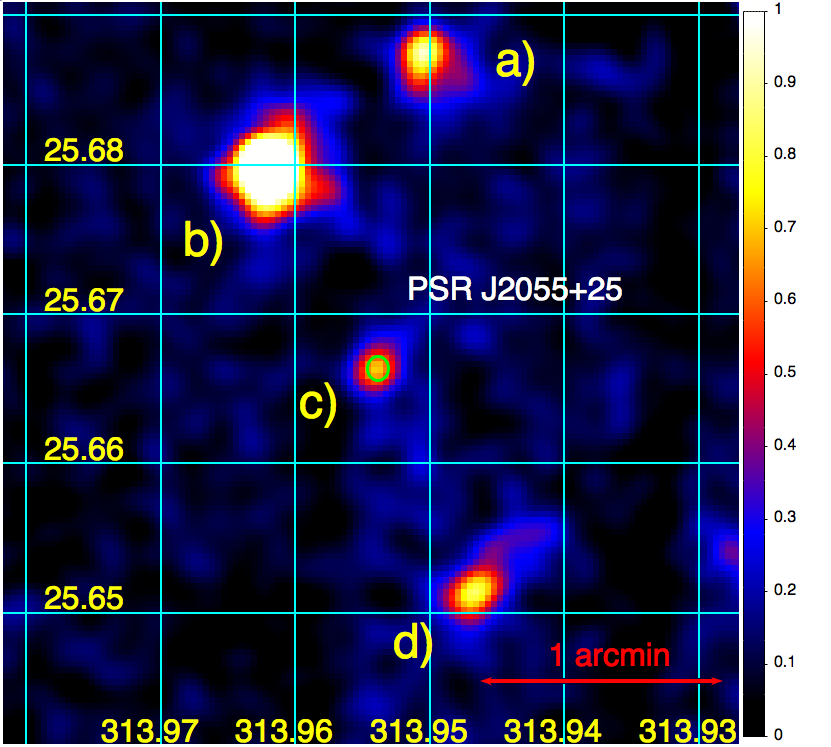}
\caption{{\it XMM-Newton} (0.2--12 keV) X-ray image of the region around PSR\,J2055+25, using $\sim$26 ks of data taken in October 2009. The color scale is in counts per square arcsecond. The green ellipse represents the current best location derived from pulsar timing, given in Table 1. North is to the top and east to the left. Four sources, labeled by letters, can be clearly identified in the image: a) XMMU J205548.0+254115, b) XMMU J205550.8+254048, c) XMMU J205549.4+253959, and d) XMMU J205547.2+253906, the ``tadpole". The best timing position of the pulsar is virtually coincident with that of XMMU J205549.4+253959. XMMU J205550.8+254048 and XMMU J205547.2+253906 are apparent in a short {\it Swift} XRT image, but the source coincident with the pulsar, XMMU J205549.4+253959, is not detected by the XRT.
\label{XMM_image}}
\end{figure}

The first five months of LAT data led to the discovery of the first 16 pulsars found from blind searches of $\gamma$-ray data~\citep{LATBSPs}. 
Here we report the discovery of an additional eight $\gamma$-ray selected pulsars by performing blind frequency searches on 11 months of 
{\it Fermi}-LAT data. These new pulsars are largely drawn from the same population as the previous 16 (see Figure~\ref{parameter_space}). Deep radio 
follow-up observations of these newly-discovered pulsars suggest that they are all either radio-quiet, or extremely radio faint.

While the less energetic $\gamma$-ray selected pulsars are found at high Galactic latitudes ($\mid$b$\mid$ $>3^\circ$), pointing to a nearby Geminga-like population, 
the remaining ones, characterized by a higher $\dot{E}$, are well aligned with the Galactic plane. Thus, the newly found pulsars represent well the 
two types of non-millisecond pulsars the LAT is unveiling: low $\dot{E}$, nearby ones at high latitude, and high $\dot{E}$, far away ones near 
the Galactic plane. Fainter, less energetic pulsars in the Galactic plane would be difficult to disentagle from the diffuse emission.

Among the low latitude $\gamma$-ray pulsars, we find PSR\,J1023--5746, a neutron star with a remarkably large rotational energy loss rate of 
$10^{37}$ erg s$^{-1}$, the largest so far of all $\gamma$-ray pulsars found in blind searches~\citep{LATBSPs}, and higher than $>$90\% of the 
entire $\gamma$-ray pulsar population~\citep{LATPulsars}. The number of such energetic gamma-ray pulsars along the Galactic plane, however, could in fact be much higher. The 
very narrow distribution in Galactic latitude of the unassociated LAT sources~\citep{LATCatalog} implies that many of them could be located at distances of several kpc, and have 
luminosities in excess of 10$^{35}$\,erg\,s$^{-1}$. Using an average value of the conversion efficiency measured for the young $\gamma$-ray pulsars detected so far~\citep{LATPulsars}, 
objects such as PSR\,J1023--5746 could account for many of the bright sources seen at low Galactic latitudes and so far unidentified.
 
\begin{acknowledgments}


The \textit{Fermi} LAT Collaboration acknowledges generous 
ongoing support from a number of agencies and institutes that have 
supported both the development and the operation of the LAT as well as 
scientific data analysis. These include the National Aeronautics and Space 
Administration and the Department of Energy in the United States, the Commissariat 
\`a l'Energie Atomique and the Centre National de la Recherche Scientifique / Institut 
National de Physique Nucl\'eaire et de Physique des Particules in France, the Agenzia 
Spaziale Italiana and the Istituto Nazionale di Fisica Nucleare in Italy, the Ministry 
of Education, Culture, Sports, Science and Technology (MEXT), High Energy Accelerator 
Research Organization (KEK) and Japan Aerospace Exploration Agency (JAXA) in Japan, and
the K.~A.~Wallenberg Foundation, the Swedish Research Council and the Swedish National Space Board in Sweden.

Additional support for science analysis during the operations phase is gratefully
acknowledged from the Istituto Nazionale di Astrofisica in Italy and the Centre National d'\'Etudes 
Spatiales in France.

Much of the work presented here was carried out on the UCSC Astronomy department's Pleiades supercomputer. 
This work made extensive use of the ATNF pulsar catalog. We thank N. Gehrels and the 
rest of the \textit{Swift} team for the \textit{Swift}/XRT observations of the LAT 
error circles of several of these newly-discovered pulsars.

The GBT is operated by the National Radio Astronomy Observatory, a facility of the National Science Foundation
operated under cooperative agreement by Associated Universities, Inc.

The Arecibo Observatory is part of the National Astronomy and Ionosphere Center, which is operated by Cornell
University under a cooperative agreement with the National Science Foundation.

The Parkes radio telescope is part of the Australia Telescope which is funded by the Commonwealth Government for operation as a National Facility 
managed by CSIRO.

\end{acknowledgments}

\bibliographystyle{apj}
\bibliography{journapj,Fermi_Bibtex_full_v2,my_bibliography}


\begin{deluxetable}{llrrrr}
\tablecolumns{6}
\tablecaption{Names and locations of the new $\gamma$-ray pulsars\label{tab:names}}
\tablewidth{0pt}
\tablehead{\colhead{PSR} & \colhead{Source Association\tablenotemark{b}}  & \colhead{R.A.\tablenotemark{c}} & \colhead{Dec.\tablenotemark{c}} &   \colhead{$l$\tablenotemark{d}} & \colhead{$b$\tablenotemark{d}}\\
	{} & {} & hh:mm:ss.s & dd:mm:ss.s & ($^\circ$) & ($^\circ$)}
\startdata
J1023--5746  &  1FGL\,J1023.0--5746 	& 10:23:02.9(5) & --57:46:05(2) 	&  284.2  &  --0.4  \\ 
	     &  0FGL\,J1024.0--5754 	& & & \\
	     &  HESS\,J1023--575 	& & & \\
	     &  CXOU\,J102302.8--574606 & & & \\
J1044--5737  &  1FGL\,J1044.5--5737	& 10:44:32.8(1)	& --57:37:19.3(8)  &  286.6  &  1.2  \\ 
	     &  1AGL\,J1043--5749 	& & & \\
J1413--6205  &  1FGL\,J1413.4--6205 	& 14:13:29.9(1) & --62:05:38(1)  &  312.4  &  --0.7  \\ 
	     &  0FGL\,J1413.1--6203 	& & & \\
	     &  1AGL\,J1412--6149 	&           &            &         &      \\
             &  EGR\,J1414--6224  	& & & \\
J1429--5911  &  1FGL\,J1429.9--5911	& 14:29:58.6(1)  	& --59:11:36.6(7) &  315.3  &  1.3  \\ 
	     &  0FGL\,J1430.5--5918 	& & & \\
J1846+0919   &  1FGL\,J1846.4+0919 	& 18:46:26.0(6)  & +09:19:46(11)  &  40.7  &  5.3  \\ 
J1954+2836   &  1FGL\,J1954.3+2836	& 19:54:19.15(4)  & +28:36:06(1)  &  65.2  &  0.4  \\ 
	     &  0FGL\,J1954.4+2838 	& & & \\
J1957+5033   &  1FGL\,J1957.6+5033 	& 19:57:38.9(8)  & +50:33:18(9)  &  84.6  &  11.0  \\ 
J2055+25\tablenotemark{a} & 1FGL\,2055.8+2539	& 20:55:48.8(2) & +25:40:02(3)  &  70.7  &  --12.5  \\ 
	     &  0FGL\,J2055.5+2540 	& & & \\
	     & XMMU\,J205549.4+253959	& & & \\
\enddata
\tablenotetext{a}{The current position uncertainty only allows for two decimal places in declination.}
\tablenotetext{b}{Sources are from the {\it Fermi} LAT First Source Catalog~\citep[1FGL, ][]{LATCatalog}, the {\it Fermi} LAT Bright Source List~\citep[0FGL, ][]{LATBSL}, EGRET~\citep[EGR, ][]{EGRcatalog}, and AGILE~\citep[1AGL, ][]{AGILEcatalog}. We also list a TeV association (HESS) and X-ray counterparts identified with {\it Chandra} (CXOU) and {\it XMM} (XMMU).}
\tablenotetext{c}{Right Ascension (R.A.) and Declination (Dec.) obtained from the timing model. The errors quoted are statistical (2.45 times the \textsc{tempo2} uncertainties). They do not account for covariance between model parameters or systematic errors caused by timing noise, which can amount to several arcseconds, and should be considered when looking for counterparts.}
\tablenotetext{d}{Galactic longitude ($l$) and latitude ($b$), rounded to the nearest decimal.}
\end{deluxetable}

\begin{deluxetable}{lllrrrrrrr}
\rotate
\tablecolumns{10}
\tablewidth{0pt}
\tablecaption{Measured and derived parameters of the $\gamma$-ray pulsars  \label{timing_solutions}}
\tablehead{\colhead{PSR} & \colhead{$n_{\gamma}$} & \colhead{$f$} &
             \colhead{$\dot{f}$} & \colhead{$\mathrm{N_{DAYS}/N_{TOAs}}$} & rms & \colhead{$\tau$} & \colhead{$\dot{E}$} & \colhead{$B_{S}$}  &\colhead{$B_{LC}$}\\
             {} & {} & \colhead{(Hz)} & {($-10^{-12}$ Hz s$^{-1}$)} & & (ms) &
             {(kyr)} & {($10^{34}$ erg s$^{-1}$)} & {($10^{12}$\,G)} & {(kG)}}
\startdata
J1023--5746  		&  4365 	&  \phn8.970827684(7)  &  30.8825(1)$^\dag$  & 550/56 & 1.0 & 4.6  &  1095.5  & 6.6&  44.0 \\ 
J1044--5737  		&  2362 	&  \phn7.192749594(2)  &  \phn2.8262(2)  & 395/21 & 0.8 &  40.3  &  80.3  & 2.8 &  9.5  \\ 
J1413--6205$^{\ddag}$  	&  5716 	&  \phn9.112389504(2)  &  \phn2.2984(3)  & 361/19 & 0.5 &  62.9  &  82.7  & 1.8 &  12.3  \\ 
J1429--5911  		&  2750 	&  \phn8.632402182(2)  &  \phn2.2728(2)  & 395/21 & 0.8 &  60.2  &  77.5  &  1.9 & 11.3  \\ 
J1846+0919  		&  1042 	&  \phn4.433578172(4)  &  \phn0.1951(5)  & 412/18 & 12.4 &  360.2  &  3.4  &  1.5 & 1.2 \\ 
J1954+2836  		&  2953 	&  10.78643292(3)  &  \phn2.4622(4)  & 383/22 & 0.9 &  69.5  &  104.8  & 1.4 & 16.4  \\ 
J1957+5033  		&  \phn449 	& \phn2.668045340(2)  &  \phn0.0504(3)  & 415/20 & 10.6 &  837.7  &  0.5  & 1.6 & 0.3  \\ 
J2055+25\phn\phn  	&  \phn715  	&  \phn3.12929129(1)  &  \phn0.040(1)  & 503/56 & 8.2 &  1226.9  &  0.5  & 1.2 & 0.3  \\
\enddata
\tablecomments{The reference epoch for the timing solutions is MJD 54800, except for PSR\,J1023--5746 (MJD 54856) and PSR\,J2055+25 (MJD 54900). $^{\ddag}$All timing solutions are valid for the 
  period MJD 54682--55016, except for that of PSR\,J1413--6205, which is only valid in the range MJD 54743--55016, due
  to a glitch which occurred around MJD 54718. Column 1 gives the
  pulsar name. Column 2 lists the number of photons obtained with the standard cuts used ({\it diffuse} class, $E >$ 300 MeV, R $< 0.8^\circ$) over the 11 month
  observational period. Columns 3 and 4 list the frequency and frequency derivative.$^\dag$PSR\,J1023--5746 requires second and third order frequency 
derivatives to model the timing noise: $\ddot{f}=7.1(2)\times10^{-21}$ Hz s$^{-2}$ and $\dddot{f}=-1.09(4)\times10^{-28}$ Hz s$^{-3}$. Column 5 gives the span of days of data 
used to generate the timing model and the number of TOAs that were generated from such data. Column 6 gives the rms timing residual of the model. Columns 7 and 8 give the characteristic
  age and spin-down luminosity. Columns 9 and 10 give the magnetic field strength at the neutron star surface and at the light cylinder. The derived quantities 
in columns 7--10 are are obtained from the timing parameters of the pulsars and are rounded to the nearest 
significant digit.}

\end{deluxetable}

\begin{deluxetable}{lccccc}
\tablecolumns{6}
\tablecaption{Pulse shape parameters for the new pulsars \label{pulse_shape_parameter_table}}
\tablehead{\colhead{PSR} & \colhead{Peak} & \colhead{$\gamma$-ray peak separation} & Reference phase  & \colhead{off-peak definition} & ROI \\
\colhead{}& \colhead{multiplicity} & \colhead{$\Delta$} & \colhead{$\phi_0$} & \colhead{$\phi$} & \colhead{($^\circ$})}

\startdata
J1023--5746  & 2  &  $0.45 \pm 0.01$ 	& 0.66 & 0.75--1.16 	& 0.8 \\ 
J1044--5737  & 2  &  $0.35 \pm 0.01$ 	& 0.62 & 0.66--1.16	& 0.8 \\ 
J1413--6205  & 2  &  $0.31 \pm 0.02$ 	& 0.85 & 0.69--1.09	& 0.8 \\ 
J1429--5911  & 2  &  $0.46 \pm 0.01$ 	& 0.86 & 0.84--1.13	& 0.8 \\ 
J1846+0919   & 1  &  \nodata 		& 0.36 & 0.50--1.09	& 0.8 \\ 
J1954+2836   & 2  &  $0.43 \pm 0.01$ 	& 0.68 & 0.84--1.16	& 0.9 \\ 
J1957+5033   & 1  &  \nodata 		& 0.77 & 0.53--1.13	& 0.8 \\ 
J2055+25     & 1  & \nodata 		& 0.83 & 0.50--1.16	& 0.8 \\
\enddata
\tablecomments{ $ $ Light curve shape parameters of each pulsar, including the peak multiplicity, the phase separation  
  between the two $\gamma$-ray peaks, the value of the phase at the reference epoch, barycentric MJD 54800 (UTC), and the
  off-peak region used in the spectral analyses. Column 6 lists the Region of Interest used to make the light curves, where we 
have selected only {\it diffuse} class events.}
\end{deluxetable}

\clearpage
\begin{deluxetable}{llllllll}
\tablecolumns{8}
\tablewidth{0pt}
\rotate
\tablecaption{Spectral parameters for the new pulsars\label{spectral_parameters}} 
\tablehead{
\colhead{PSR} & \colhead{Photon Flux ($F_{100}$)} & \colhead{Energy Flux ($G_{100}$)} & \colhead{$\Gamma$}  
& \colhead{$E_{\rm{cutoff}}$}  & \colhead{$TS$}  & \colhead{$TS_{\rm{cutoff}}$} & \colhead{$d_{ps}$}\\
\colhead{} & \colhead{($\rm 10^{-8} \,ph \ cm^{-2} \,s^{-1}$)} & \colhead{($\rm 10^{-11} \,erg \ cm^{-2}\, s^{-1}$)} 
& \colhead{} & \colhead{($\rm GeV$)}  & \colhead{}  & \colhead{} & \colhead{(kpc)}}

\startdata
J1023--5746 	&  41.5 $\pm$ 0.5 & 26.9 $\pm$ 1.8 &  1.58 $\pm$ 0.13	& 1.8 $\pm$ 0.3	& 686  & 88 & 2.4 \\
J1044--5737  	&  14.3 $\pm$ 1.7 & 10.3 $\pm$ 0.65 &  1.60 $\pm$ 0.12	& 2.5 $\pm$ 0.5 & 799  & 69.7 & 1.5 \\
J1413--6205  	&  12.9 $\pm$ 2.2 & 12.9 $\pm$ 1.0  &  1.32 $\pm$ 0.16	& 2.6 $\pm$ 0.6 & 461  & 64.9 & 1.4\\
J1429--5911  	&  16.2 $\pm$ 2.4 & 9.26 $\pm$ 0.81 &  1.93 $\pm$ 0.14	& 3.3 $\pm$ 1.0 & 318  & 20.8 & 1.6\\
J1846+0919  	&  \phn4.1 $\pm$ 0.77 & 3.58 $\pm$ 0.35 &  1.60 $\pm$ 0.19	& 4.1 $\pm$ 1.5 & 363  & 19.8 & 1.2\\
J1954+2836  	&  11.9 $\pm$ 1.6 & 9.75 $\pm$ 0.68 &  1.55 $\pm$ 0.14	& 2.9 $\pm$ 0.7 & 595  & 54.3 & 1.7\\
J1957+5033  	&  \phn3.3 $\pm$ 0.52 & 2.27 $\pm$ 0.20 &  1.12 $\pm$ 0.28	& 0.9 $\pm$ 0.2 & 395  & 39.3 & 0.9\\
J2055+25  	&  11.1 $\pm$ 1.1 & 11.5 $\pm$ 0.70 &  0.71 $\pm$ 0.19	& 1.0 $\pm$ 0.2 & 779  & 127 & 0.4\\
\enddata

\tablecomments{Results of the unbinned maximum likelihood spectral fits for the new $\gamma$-ray pulsars (see Section~\ref{spectral}). 
Columns 2 and 3 list the on-peak (defined as the complement of the off-peak region defined in column 5 of Table~\ref{pulse_shape_parameter_table}) photon flux $F_{\rm 100}$ 
and on-peak energy flux $G_{\rm 100}$ respectively. 
The fits used an exponentially cutoff power law model (see Equation 1) with photon index $\Gamma$ and cutoff energy $E_{\rm{cutoff}}$ 
given in columns 4 and 5. The systematic uncertainties on $F_{\rm 100}$, $G_{\rm 100}$, and $\Gamma$ due to uncertainties in the 
Galactic diffuse emission model have been added in quadrature with the statistical errors. Uncertainties in the instrument response 
induce additional biases of $\delta F_{100} = (+30\%,\,-10\%)$, $\delta G_{100} = (+20\%,\,-10\%)$, $\delta\Gamma = (+0.3,\,-0.1)$, 
and $\delta E_{\rm{cutoff}} = (+20\%,\,-10\%)$. The test statistic ($TS$) for the source significance is provided in column 6. The 
significance of an exponential cutoff (as compared to a simple power law) is indicated by $TS_{\rm cutoff}$ in column 7, where a value 
$> 10$ indicates that the exponentially cutoff model is significantly preferred in every case. Column 8 gives a pseudo-distance estimate, $d_{ps}$, calculated using 
equation (2), described in Section~\ref{distances}.}
\end{deluxetable}
\normalsize
\noindent

\clearpage

\begin{deluxetable}{lllrrrrrrr}
\tabletypesize{\footnotesize}
\rotate
\tablewidth{0pt}
\tablecaption{Radio observations of $\gamma$-ray selected pulsars\label{tab:radiolims}}
\tablehead{
\colhead{Target} & \colhead{Obs Code} & \colhead{Date} & \colhead{$t_\mathrm{obs}$} & \colhead{DM max\tablenotemark{a}} &\colhead{R.A.\tablenotemark{b}} & \colhead{Decl.\tablenotemark{b}} & \colhead{Offset\tablenotemark{c}} & \colhead{$T_\mathrm{sky}$\tablenotemark{d}} & \colhead{$S_\mathrm{min}$\tablenotemark{e}} \\
 & & & {(s)} & (pc\,cm$^{-3})$ & {(J2000)} & {(J2000)} & {(arcmin)} & {(K)} & {($\mu$Jy)}}
\startdata
J1023--5746 & Parkes-AFB   & 2009 Apr 14 & 16200 & 2000 & 10:23:10.6  & --57:44:20 & 2.0 & 6.0  & 31 \\
            & Parkes-10cm  & 2009 Nov 26 & 11500 & 1000 & 10:23:02.3  & --57:46:09 & 0.1 & 0.9  & 22 \\
J1044--5737 & Parkes-BPSR  & 2009 Aug 02 & 16200 & 2000 & 10:44:33.3  & --57:37:15 & 0.1 & 4.2  & 21 \\
J1413--6205 & Parkes-BPSR  & 2009 Aug 02 & 16200 & 2000 & 14:13:14.2  & --62:04:34 & 2.1 & 8.1  & 25 \\
J1429--5911 & Parkes-BPSR  & 2009 Aug 03 & 16200 & 2000 & 14:30:02.2  & --59:11:20 & 0.5 & 6.5  & 22 \\
J1846+0919  & AO-Lwide     & 2009 Aug 07 & 3600  & 2000 & 18:46:26.8  & +09:19:42  & 0.2 & 3.8  & 4  \\
J1954+2836  & AO-Lwide     & 2009 Aug 04 & 1200  & 1310 & 19:54:18.9  & +28:36:10  & 0.1 & 3.2  & 7  \\
            & AO-Lwide     & 2009 Oct 14 & 2700  & 1310 & 19:54:19.2  & +28:36:06  & 0.0 & 3.2  & 4  \\
J1957+5033  & GBT-820      & 2009 Sep 21 & 4900  & 1100 & 19:57:32.3  & +50:36:33  & 3.4 & 6.9  & 25 \\
J2055+25    & AO-327       & 2009 May 15 & 1800  & 2500 & 20:55:34.8  & +25:40:23  & 3.2 & 62.2 & 85  \\
            & AO-Lwide     & 2009 May 15 & 1800  & 2500 & 20:55:34.8  & +25:40:23  & 3.2 &  1.4 & 106 \\
\enddata
\tablenotetext{a}{Maximum value of dispersion measure (DM) searched.}
\tablenotetext{b}{Telescope pointing direction (not necessarily source position).}
\tablenotetext{c}{Offset between telescope pointing direction and pulsar position.}
\tablenotetext{d}{Sky temperature estimated by scaling 408 MHz~\citep{haslam81} all sky map to observing frequency using spectral index of $-2.6$.}
\tablenotetext{e}{Flux limits are at the observing frequency, not scaled to an equivalent 1.4 GHz flux.}
\end{deluxetable}

\begin{deluxetable}{llrrrrrrr}
\tablewidth{0pt}
\tabletypesize{\small}
\tablecaption{Definition of observing codes\label{tab:radioobs}\tablenotemark{a}}
\tablehead{
\colhead{Obs Code} & \colhead{Telescope} & \colhead{Gain} & \colhead{Freq} & \colhead{$\Delta f$} & \colhead{$\beta$\tablenotemark{b}} & \colhead{$n_\mathrm{p}$} & \colhead{HWHM} & \colhead{$T_\mathrm{rec}$} \\
 & & (K/Jy) & (MHz) & (MHz) &   &  & (arcmin) & (K)
}
\startdata
GBT-820      & GBT     & 2.0  & 820  & 200 & 1.05 & 2  & 7.9 & 29 \\
AO-327       & Arecibo & 11   & 327  & 50  & 1.12 & 2  & 6.3 & 116 \\
AO-Lwide     & Arecibo & 10   & 1510 & 300 & 1.12 & 2  & 1.5 & 27 \\
Parkes-AFB   & Parkes  & 0.735& 1374 & 288 & 1.25 & 2  & 7.0 & 25 \\
Parkes-BPSR  & Parkes  & 0.735& 1352 & 340 & 1.05 & 2  & 7.0 & 25 \\
Parkes-10cm  & Parkes  & 0.67 & 3078 & 864 & 1.25 & 2  & 3.3 & 30 \\

\enddata
\tablenotetext{a}{The parameters in this table refer to Equation 3 in the text.}
\tablenotetext{b}{Instrument-dependent sensitivity degradation factor.}
\end{deluxetable}

\clearpage

\end{document}